\documentclass[aps,pra,graphicx,9pt,twocolumn,superscriptaddress]{revtex4-2}
\usepackage{amsmath,amscd,graphicx,graphicx,cases}
\usepackage{multirow,float,subfigure,appendix,bm}
\usepackage{diagbox,makecell,blindtext,color,amssymb}
\usepackage[colorlinks,linkcolor=blue,urlcolor=blue,
anchorcolor=blue,citecolor=blue]{hyperref}
\usepackage{textcomp,booktabs,colortbl,easyReview}
\definecolor{mygray}{gray}{.9}
\newcolumntype{C}{c<{\kern\tabcolsep}@{}}

\begin{document}
\title{Thermal-dephasing-tolerant generation of mesoscopic superposition states with Rydberg dressed blockade}
\author{Ri-Hua Zheng}
\affiliation{Fujian Key Laboratory of Quantum Information and Quantum Optics, College of Physics and Information Engineering, Fuzhou University, Fuzhou, Fujian 350108, China}
\author{S.-L. Su}\email{slsu@zzu.edu.cn}
\affiliation{School of Physics, Zhengzhou University, Zhengzhou 450001, China}
\author{Jie Song}
\affiliation{Department of Physics, Harbin Institute of Technology, Harbin 150001, China}
\author{Weibin Li}\email{weibin.Li@nottingham.ac.uk}
\affiliation{School of Physics and Astronomy, University of Nottingham, Nottingham NG7 2RD, United Kingdom}
\author{Yan Xia}\email{xia-208@163.com}
\affiliation{Fujian Key Laboratory of Quantum Information and Quantum Optics, College of Physics and Information Engineering, Fuzhou University, Fuzhou, Fujian 350108, China}
\begin{abstract}
Multipartite entangled states involving non-locality are one of the most fascinating characteristics of quantum mechanics. 
In this work, we propose a thermal-dephasing-tolerant generation of mesoscopic entangled states with Rydberg dressed atoms. 
We encode logical state on dressed states rather than Rydberg states.
Such treatment can increase the lifetime of multipartite  entanglement coherence to around 3 times compared to the Rydberg-state-coding one at the same system size, and therefore induce solid fidelities of mesoscopic superposition states generation. 
The current work theoretically verifies the advantages of using Rydberg dressed states in many-body quantum entanglement, which is helpful for large-scale quantum computation and many-body Rydberg quantum simulation.

\end{abstract}	
\maketitle
\section{introduction}
Rydberg atoms exhibit very strong van der Waals interactions \cite{Saffman2010RMP}, which can be flexibly modulated by varying the interatomic distances and therefore show great potential in the quantum simulations of many-body systems \cite{Schaus2012Nature,Hofmann2013PRL,lukin-256,Bluvstein2022Nature}.
Many applications based on Rydberg atoms rely on the ability to coherently manipulate atoms on time scales below the radiation lifetime of the excited state.
Interestingly, it has been found \cite{Henkel2010PRL,dressed-PRL,Molmer-dressedGB-PRA} that weakly admixing excited Rydberg states with laser light can extend this time scale limitation.
This proposal can also be utilized to produce a novel type of long-range interactions between Rydberg-dressed ground state atoms \cite{Johnson-dressedGB-PRA,Balewski-dressedGB-NJP}, which provides a tool to enhance coherence of many-body systems from inelastic collisions and spontaneous emission, such as the production of exotic quantum phases \cite{Johnson-dressedGB-PRA,dressed-PRL,dressed-Science}, spin squeezing \cite{Molmer-dressedGB-PRA,dressedspin}, quantum computation \cite{dressedQC1,dressedQC2}, and entanglement generation \cite{dress_cat_1,dress_cat_2,Jau-GB-NP,shao-GB,shao-GB-GHZ3}.


Multipartite entanglement lies at the heart of many-body quantum simulation. 
Among diverse types of entangled states, the Greenberger-Horne-Zeilinger (GHZ) states \cite{GHZ} (or called  two-component atomic cat states \cite{atomcat,atomcat2}), have attracted a lot of research interest \cite{cat_interest_1,cat_interest_2,cat_interest_3} because of their characteristics of the maximum entanglement. 
Recently, two experimental works \cite{lukinScience,songcat} simultaneously report the generation of 20-qubit GHZ states with similar fidelity $\mathcal{F}\sim0.55$ on the Rydberg atoms platform and the superconducting qubits platform, respectively.
One of the major obstacles to achieving higher fidelity for generating multipartite entanglement \cite{lukinScience}, as well as other extensive quantum simulation and quantum computation based on Rydberg atoms \cite{Forster,highfideliyNP,kang1,wangyu,su1,zheng1,kang2,su2,liushuai}, is thermal dephasing. 
Due to the finite temperature of neutral atoms, the Doppler shift occurs on each site of the Rydberg atom arrays \cite{lukinScience,Dopplerdata-PRL}.
Additionally, the thermal noise causes fluctuations of atomic distance (for example, $\delta R/R\sim 6\%$ \cite{Forster}), i.e., disturbing the desired electric dipole-dipole interaction (EDDI) strength. 
The above two disturbances together constitute the thermal dephasing mechanisms, which have hindered the high-precision completion of many quantum works \cite{Forster,lukinScience,highfideliyNP} with Rydberg atoms.
Especially when multiple Rydberg atoms are excited to Rydberg states, the thermal dephasing, quantified by the lifetime of coherence, $T_2$, becomes shorter with the increase in the number of atoms $N$ (specifically, $T_2\sim$0.6 $\mu$s for $N=20$ \cite{lukinScience}). 
Therefore, Rydberg dressing is a promising candidate to solve these problems caused by thermal dephasing.

In this work, we propose to generate mesoscopic superposition states through a Rydberg dressing induced blockade. 
We dress the Rydberg atoms probability $P_r=25\%$, leading to coherence lifetimes $T_2=$\{11.0, 8.2, 7.2, 6.2, 5.7, 5.1\} $\mu$s for \{4, 6, 8, 10, 12, 14\}-atom GHZ states, 
around 3 times of $T_2$ in the work of \cite{lukinScience}, at the same system size.
For larger scale systems, the dressed lifetime $T_2$ decreases with the system size $N$ as $1/\sqrt{N}$, which is the same as the scaling rule in the work without Rydberg dressing \cite{lukinScience}.
The fidelities are \{97.2\%, 95.1\%, 91.1\%, 84.6\%\} for \{4, 6, 8, 10\}-atom GHZ states when considering the thermal dephasing.
Based on the above fidelities data, we give a prediction fidelity of 80\% for the 20-atom GHZ state.
The balance between dressing low-percentage Rydberg states and shortening the evolution time in large-scale neutral atoms systems in the present work may provide a reference for quantum computation and quantum simulation based on multiple dressed atoms.
The Rydberg-dressing method for many-body entanglement generation is scalable and dephasing-tolerant, which will contribute to realizing near-term quantum simulation and quantum computation with Rydberg atom arrays,
such as  entanglement-enhanced sensing \cite{QS-GHZ}, quantum metrology \cite{QM-GHZ}, and  quantum error correction \cite{QEC-GHZ} based on GHZ states.

The article is organized as follows.
In Sec. \ref{Physical Model}, we give the physical model for the generation of multipartite entanglements with the Rydberg dressing. 
In Sec. \ref{Thermal dephasing}, we discuss the main causes of the thermal dephasing in Rydberg atoms. 
Scaling of fidelities and $T_2$ are placed in Secs. \ref{Scaling_F} and \ref{Scaling_T2}, respectively. 
The conclusion is given in Sec. \ref{Conclusion}.

\begin{figure}
\includegraphics[width=8.4cm]{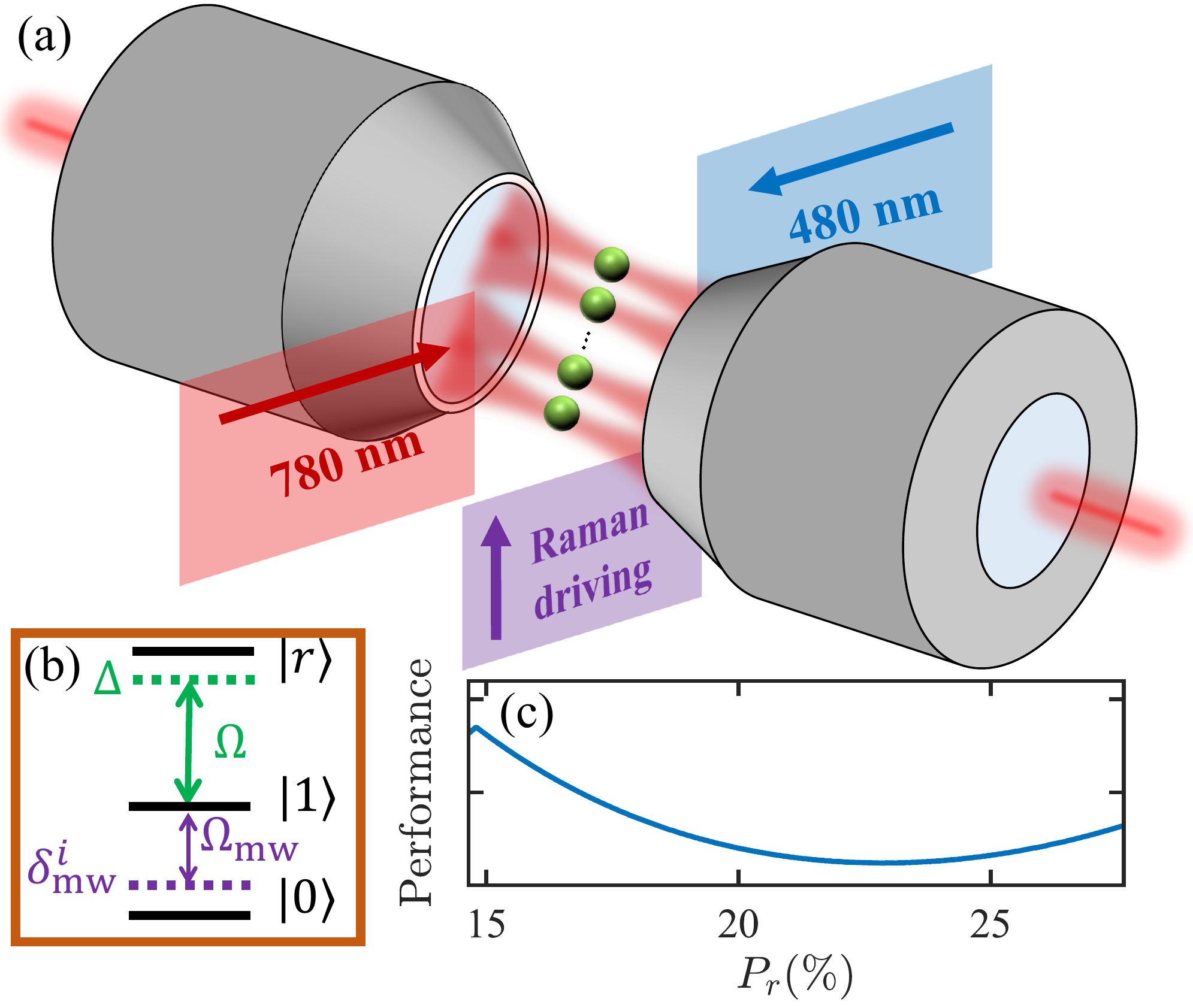}
\caption{(a) Envisioned experimental setup. 
All atoms fixed on the one-dimensional array are driven by Rydberg dressed lasers (780 and 480 nm) and Raman (microwave) drivings with the same detunings and Rabi frequencies. 
Two addressing lasers (not shown) are added on the two edge atoms to differentiate microwave detunings from other atomic ones.
(b) Energy levels and couplings of each atom. 
(c) Performance of the entanglement generation versus dressed percentage $P_r$.
This performance index combines the impact of the lifetime, dressed interaction, and dressed interaction bandwidth (See the text for details).}
\label{model}
\end{figure} 

\section{Physical Model}\label{Physical Model}
Consider a one-dimensional array with an even number $N$ of neutral $^{87}{\rm Rb}$ atoms, trapped in the optical tweezers, as shown in Fig. \ref{model}. For all the atoms, we encode  two ground states $|0\rangle=|5S_{1/2},F=1,m_F=1\rangle$ and $|1\rangle=|5S_{1/2},F=2,m_F=2\rangle$ and one excited Rydberg state $|r\rangle=|70S,J=1/2,m_J=-1/2\rangle$ \cite{lukinPRL,lukinScience}.
The coupling between $|1\rangle$ and $ |r\rangle$ is driven by a two-photon transition with effective coupling strength $\Omega$ and detuning $\Delta$.
Additionally, the microwave-frequency fields with strength $\Omega_{\rm mw}(t)$ and detuning $\delta_{\rm mw}^n(t)$ are added for driving the rotations between the ground states $|0\rangle$ and $|1\rangle$. 
We choose the nearest-neighbor EDDI energy, $V/(2\pi)=21$ MHz, i.e., equal adjacent atomic intervals of 5.87 $\mu$m and $C_6=858$ GHz $\mu {\rm m}^6\cdot h$. 
The Hamiltonian here is given by ($\hbar=1$ and $n,i,j\le N$)
\begin{subequations} \label{H}
\begin{align}
H&=H_{\rm RDB}+H_{\rm mw}, \\
H_{\rm RDB}&=\sum_{n=1}^N\frac{\Omega}{2}\sigma_{x,r1}^n +\Delta \sigma_{rr}^n+\sum_{i<j}\frac{V}{|i-j|^6}\sigma_{rr}^i \sigma_{rr}^j, \\
H_{\rm mw}&=\sum_{n=1}^N[\Omega_{\rm mw}(t)\sigma_{x,0d}^n  +[U_1-\delta_{\rm mw}^n(t)]\sigma_{00}^n,
\end{align}
\end{subequations}
with $\sigma^n_{\alpha\alpha}=|\alpha\rangle_n\langle \alpha|$,
$\sigma^n_{x,\alpha \beta}=|\alpha\rangle_n\langle \beta|+|\beta\rangle_n\langle \alpha|$
$(\alpha=r,0; \ \beta=1,d)$,
where $|d\rangle$ is the dressed state, given by $|d\rangle=-{\rm sign}(\Delta) \sin (\theta/2)|1\rangle+\cos(\theta/2)|r\rangle$ with $\sin \theta=\Omega/\sqrt{\Omega^2+\Delta^2}$ and $\cos \theta=-{\rm sign}(\Delta) \Delta/\sqrt{\Omega^2+\Delta^2}$.
Therefore, the percentage $P_r$ of Rydberg states in the dressed states is $\cos^2 (\theta/2)$.
Note the driving between $|0\rangle$ coupling $|r\rangle$ is also a two-photon transition.
For each pair of adjacent atoms, after abandoning decoupled anti-symmetric state $(|1r\rangle-|r1\rangle)/\sqrt{2}$,
the Hamiltonian can be reduced to 
\begin{eqnarray}\label{dressedH}
H_{\rm RDB}^{(2)}=\left(
\begin{array}{ccc}
0 &\frac{\Omega}{\sqrt{2}} & 0 \\
\frac{\Omega}{\sqrt{2}} &\Delta & \frac{\Omega}{\sqrt{2}} \\
0 &\frac{\Omega}{\sqrt{2}} & 2\Delta+V \\
\end{array}
\right),
\end{eqnarray}
in basis \{$|11\rangle$, $(|1r\rangle+|r1\rangle)/\sqrt{2}$, $|rr\rangle$\}. 
Equation \ref{dressedH} is insightful in that it allows us to understand the energy scales directly. 

\begin{figure}
\includegraphics[width=8cm]{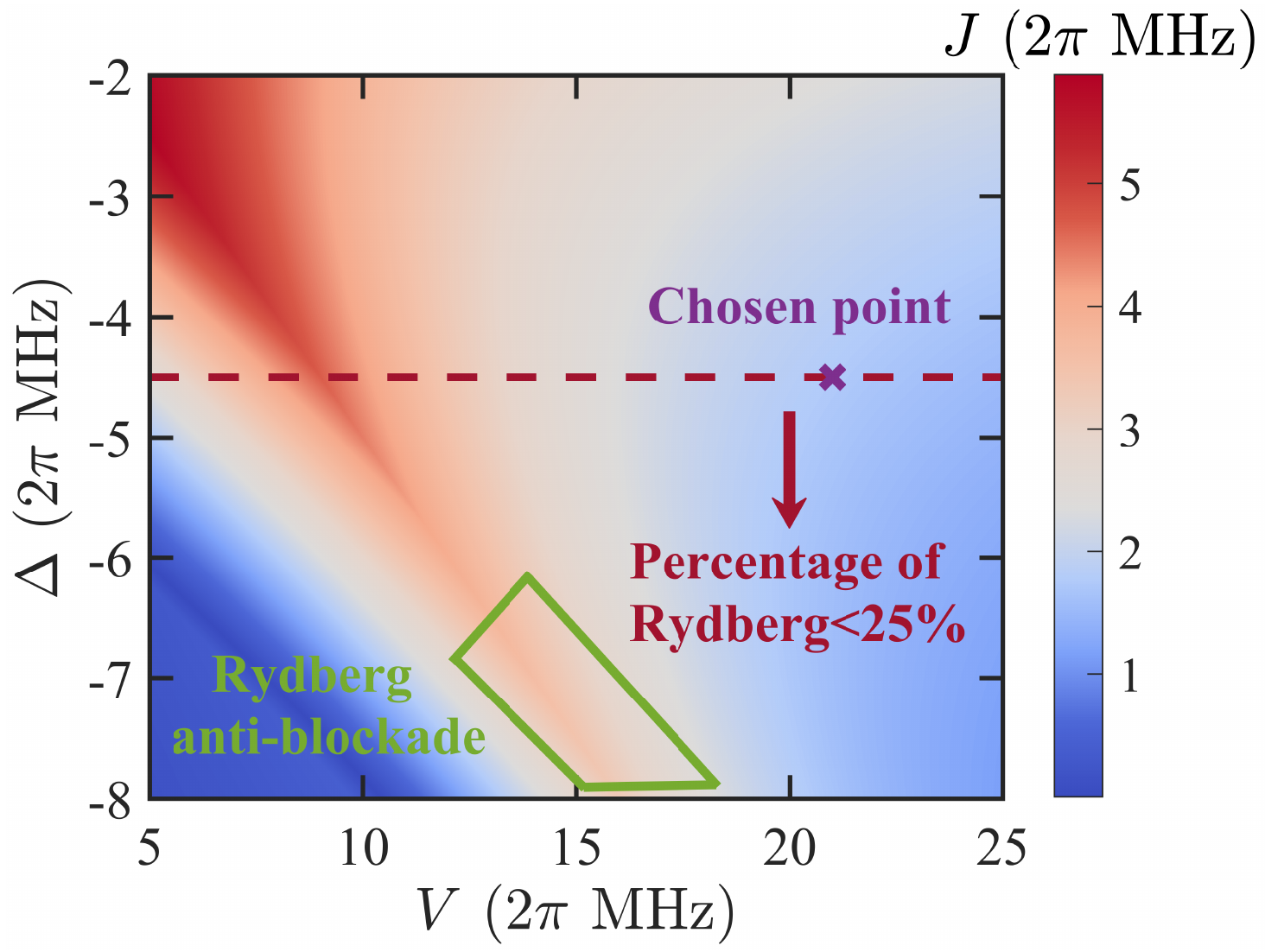}
\caption{(Color online) Dressed energy $J$ versus detuning $\Delta$ and EDDI strength $V$. 
The effective coupling strength $\Omega$ is fixed at $2\pi\times 8$ MHz.
The area where the percentage of Rydberg states $P_r$ is less than 25\%, is marked below the red dotted line. 
The stick-shaped area represents the Rydberg anti-blockade with weak regimes (not satisfying $\Delta\gg\Omega$).
The selected point $(V,\Delta)/(2\pi)=(21,-4.5)$ MHz is marked by a cross.}
\label{dressedJ}
\end{figure}

The performance of the entanglement generation is quantified as ${\rm per.}= J \cdot T_2 \cdot {\rm bw}_J$, shown in Fig. \ref{model}(c) with free units,
where $J$ and ${\rm bw}_J$ are the dressed energy and dressed energy bandwidth, respectively.
The dressed energy, arising from the Stark shifts caused by the reciprocity among the EDDI and dressed lasers drivings, is given by $J=|2U_1-U_2|$, with $U_1=(\Delta-{\rm sign}(\Delta)\sqrt{\Omega^2+\Delta^2})/2$ and $U_2=-{\rm sign}(\Delta) \cdot \min|{\rm eig}(H_{\rm RDB}^{(2)})|$  representing single- and double-atom Stark shifts on the dressed states, respectively. 
When the Rydberg blockade effect is strong enough ($V\gg\{\Omega, \Delta\}$), one can calculate that $U_2 = (\Delta-{\rm sign}(\Delta)\sqrt{2\Omega^2+\Delta^2})/2$, which is coincide with the results in \cite{Johnson-dressedGB-PRA,Jau-GB-NP,Balewski-dressedGB-NJP}. For illumination, we plot the dressed energy $J$ versus detuning $\Delta$ and EDDI strength $V$ in Fig. \ref{dressedJ}. 

As exhibited in Fig. \ref{dressedJ}, the stick-shaped area around the green dashed line processes higher dressed energy $J$. 
This area demonstrates an interesting physical phenomenon, called the Rydberg anti-blockade \cite{CTPRL98,TCPRL104,ZKPRA82,THPRL108,WCPRL110,SEPRA93}.
In principle, in all the regions of Fig. \ref{dressedJ}, the dressed energy $J$ on the two-atom dressed states $|dd\rangle_{i(i+1)}$ blockade the associated transitions from the ground state to $|dd\rangle_{i(i+1)}$ [adjusting $J\gg\Omega_{\rm mw}$, dropping $(t)$ hereafter]. 
That is another phenomenon with rich physical connotations, called the Rydberg dressed blockade (RDB) \cite{Molmer-dressedGB-PRA,Johnson-dressedGB-PRA,Balewski-dressedGB-NJP,Jau-GB-NP,shao-GB,shao-GB-GHZ3}, which acts like the Rydberg blockade, however with a difference that the transitions to adjacent atoms in the dressed state $|dd\rangle_{i(i+1)}$ rather than the two-atom Rydberg states $|rr\rangle_{i(i+1)}$ are forbidden. 

To shorten the evolution time of the dynamics induced by $H_{\rm mw}$, we prefer a larger dressed energy $J$ for RDB. 
On the other hand, to reduce the thermal dephasing of atoms, the composition of Rydberg states $P_r$ should be small (resisting atomic Doppler shifts).
Additionally, the dressed energy $J$ needs to be stable within a certain fluctuation range of $V$ (resisting atomic position floating). 
Ergo, we choose $(V,\Delta)/(2\pi)=(21,-4.5) \ {\rm MHz}$ with $P_r=25\%$.

Up to now, the light shift (caused by the dressed laser) for each atom can be canceled by the $U_1 |0\rangle_n \langle 0|$ term of the Hamiltonian of microwave-frequency fields, $H_{\rm mw}$. 
Then we use $H_{\rm mw}$ to flexibly manipulate the dynamics of multiple atoms in subspace $\mathcal{S}_1=$\{$|0\rangle$, $|d\rangle$\} (equal to an $N$-qubit quantum system). 
Moreover, due to the RDB effect, the subspace $\mathcal{S}_2=\{|dd\rangle_{i(i+1)}\}$ are forbidden. 
The calculation space is further locked into $\mathcal{S}_3=\complement_{\mathcal{S}_1}\mathcal{S}_2$.
Therefore we successfully reduce the calculation dimension of the $N$-qubit quantum system from $2^N$ to $\sum_{m=0}^{N/2}C_{N+1-m}^m$ (see Appendix \ref{Appendix_A} for details), analogous to the multiple-qubit work by Omran $et$ $al.$ \cite{lukinScience}. 
Remarkably, such a reduction of the dimension of the system subspace induces a non-local effect to generate entanglement \cite{Jau-GB-NP,lukinScience}. 
Since the dynamics of the full Hilbert space and the reduced space are not completely equivalent, we use the projection ratio ${{\cal P}_N={\rm Tr}_{{\cal S}_2}\rho_f}$ of the $N$-atom reduced subspace relative to the corresponding full Hilbert space as a benchmark for the divergence between them. For the systems with ${N=4,6,8,10,12}$, we obtain ${{\cal P}_N=0.98,0.95,0.95,0.93,0.90}$.

Further, by appropriately adjusting the addressing drivings applied on edge atoms, we can make their microwave detunings ($\delta_{\rm mw}^e$) different from other atomic ones ($\delta_{\rm mw}^{ne}$), 
i.e., $\delta_{\rm mw}^1=\delta_{\rm mw}^N=\delta_{\rm mw}^e\neq\delta_{\rm mw}^p=\delta_{\rm mw}^{ne}$ ($p=2,3,...,N-1$). 
When $\delta_{\rm mw}^e$ is closed to $\delta_{\rm mw}^{ne}$ but differs in certain values, the initial state $|0000\cdots\rangle$ will be driven to the GHZ state $| {\rm GHZ}_N\rangle=(|d0d0\cdots d0\rangle+|0d0d\cdots0d\rangle)/\sqrt{2}$ by adiabatically increasing detuning ($-\delta_{\rm mw}^{n}$) from negative large values to positive large values (see  Appendix \ref{Appendix_B} for deviations). 
Such an adiabatic method has been proved in the work of Ref. \cite{lukinScience}, and they optimize their pulses through a remote dressed chopped-random basis algorithm \cite{OPT1,OPT2}. 
Here we utilize another effective optimal control method, gradient ascent pulse engineering (GRAPE) \cite{GRAPE} (recently proved to be a precise algorithm in the experiment \cite{sunGRAPE1,sunGRAPE2,sunGRAPE3,sunGRAPE4}), to shorten the evolution time of the above adiabatic process for the multipartite GHZ states generation. 
The optimization results are exhibited in Table \ref{GRAPEtable}. 
Note that the maximum absolute value of $\Omega_{\rm mw}$ does not exceed $2\pi \times 0.14$ MHz to ensure the stability of the RDB dynamics ($J\gg\Omega_{\rm mw}$). 
Detailed driving pulses for \{4, 6, 8, 10\}-atom systems can be seen in Appendix \ref{Appendix_C}.

\begin{table}
\centering
\begin{tabular}{@{}CCc@{}}
\toprule
Number of atoms & \ \ Time ($\mu$s) \ \  & Ideal fidelity \\
\midrule
4 & 2.1 & 0.99 \\
\rowcolor{mygray}
6 & 3.1 & 0.99\\
8 & 4.2 & 0.99 \\
\rowcolor{mygray}
10 & 5.3 & 0.99 \\
12 & 5.9 & 0.99 \\
\rowcolor{mygray}
14 & 7.6 & 0.99 \\
16 & 8.3 & 0.99 \\
\bottomrule
\end{tabular}
\caption{Ideal fidelities and preparation times optimized by the GRAPE.
The ideal fidelity is calculated in the effective subspace $\mathcal{S}_3$ (no $\{|dd\rangle_{i(i+1)}\}$ subspace).}
\label{GRAPEtable}
\end{table}
\section{Thermal dephasing}\label{Thermal dephasing}
We next consider a very critical experimental imperfection, the thermal dephasing noise, in the numerical simulation.
The finite temperature of atoms induces their nonzero spread velocity and position uncertainty. 
Such that the thermal dephasing mechanism is determined by these two factors. 

(i) The position uncertainty can be modeled by the Gaussian distribution with the probability density function being $\mathcal{N}(x_n,\sigma^2)=e^{-(x-x_n)^2/(2\sigma^2)}/(\sqrt{2\pi}\sigma)$, where $x_n$ the ideal position of the $n$th atom and $\sigma^2$ the variance. 
The adjacent distance of the atoms is $R=\mathcal{N}(x_i,\sigma^2)-\mathcal{N}(x_{i+1},\sigma^2)=\mathcal{N}(R_0,2\sigma^2)$ with $R_0=x_i-x_{i+1}=5.87 \ \mu$m. Specifically, the standard deviation $\sigma$ of Gaussian distribution for an atom in a finite temperature $\sim$10 $\mu$K is around 0.1 $\mu$m \cite{lukinScience,Ahmed-letter}. As $V=C_6/R^6$, a spread in interaction strength can be calculated by the position distribution and further simulated with the original Hamiltonian in Eq. (\ref{H}). 

(ii) The spread velocity causes fluctuating Doppler shifts.
According to concrete reports \cite{Dopplerdata-PRL,dopp-PRA,Wu-Dopp}, the Doppler shift leads to a random detuning $\delta^D$ on each atom-array site, viz., $\delta^D\in \mathcal{N}(0,\sigma_D^2)$, simulated as an error term $H^D=\delta^D |r\rangle \langle r|$ added to the original Hamiltonian in Eq. (\ref{H}).
The value of $\sigma^D$ is given by $\sigma^D=k_{\rm eff}\Delta v$ with $k_{\rm eff}=|\boldsymbol{k_{|1\rangle \leftrightarrow |e\rangle}}+\boldsymbol{k_{|e\rangle \leftrightarrow |r\rangle}}|$ the effective wave vector of two-photon transition $|1\rangle \leftrightarrow |r\rangle$ and $\Delta v=\sqrt{k_BT/m}$ the one-dimensional root-mean-square (rms) velocity spread of atoms.
Bringing in the values of Boltzmann constant $k_B$, atomic temperature $T=10$ $\mu$K \cite{lukinScience}, and atomic mass $m=87\times 1.66 \times10^{-27}$ kg, 
we can give the rms velocity spread $\Delta v=\sqrt{k_BT/m}=0.031$ m/s.
On the other hand, one can drive $|1\rangle \leftrightarrow |e\rangle$ ($|e\rangle \leftrightarrow |r\rangle$) by a 780 nm (480 nm) laser with 
$|\boldsymbol{k_{|1\rangle \leftrightarrow |e\rangle}}|=2\pi/780 \ {\rm nm}^{-1}$ 
($|\boldsymbol{k_{|e\rangle \leftrightarrow |r\rangle}}|=2\pi/480 \ {\rm nm}^{-1}$).
These two lasers can be focused on the atom array by opposite directions \cite{Dopplerdata-PRL} to minimize the Doppler shifts, resulting in $k_{\rm eff}=2\pi/480 \ {\rm nm}^{-1}-2\pi/780 \ {\rm nm}^{-1}$ and further $\sigma_{D}/(2\pi)=24.77$ kHz.
Since the energy of ground states $|0\rangle$ and $|1\rangle$ is closed, the Doppler shifts are almost the same for $|1\rangle \leftrightarrow |r\rangle$ and $|0\rangle \leftrightarrow |r\rangle$. 
Therefore the error term $H^D=\delta^D |r\rangle \langle r|$ can well describe the Doppler shifts effect in the present atomic systems.

\section{Scaling of fidelities}\label{Scaling_F}
\begin{figure}
\includegraphics[width=8cm]{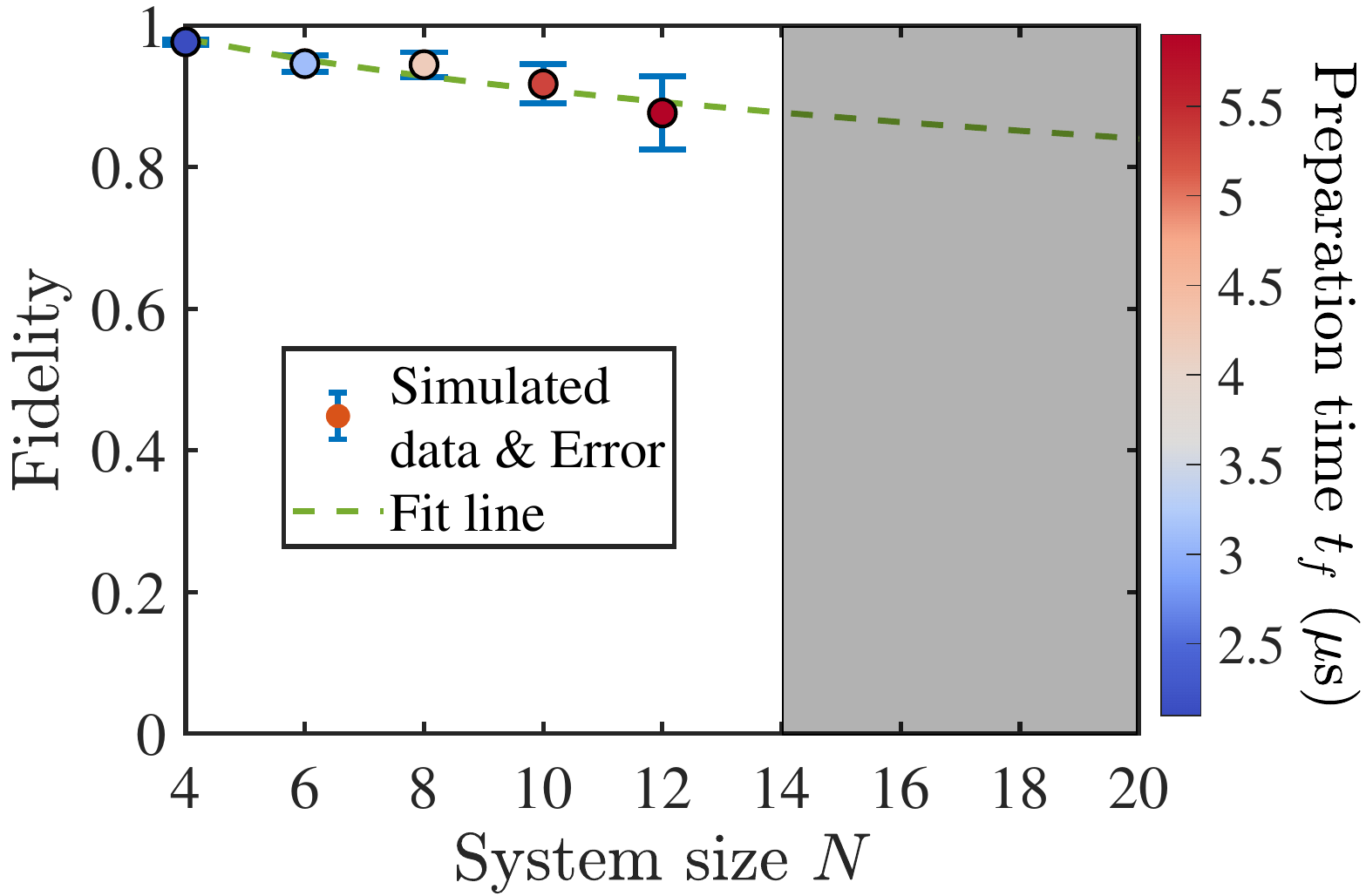}
\caption{Fidelities of GHZ states in the systems at different scales. 
The colors of dots scale the corresponding preparation time for 
GHZ states.
The green dashed line exhibits the fitting results [based on $\ln (\mathcal{F}-1/2) \propto -\sqrt{N}$] induced from the fidelities for \{4, 6, 8, 10, 12\}-atom systems.
This line gives a prediction of the fidelities for larger $N\ge 14$ and also more difficult to compute systems, marked by a gray area.
}
\label{Fidelity}
\end{figure}

Considering the above two negative factors, the simulation results for GHZ states preparation in \{4, 6, 8, 10\}-atom systems are shown in Fig. \ref{Fidelity}. 
The full Hamiltonian in Eq. (\ref{H}) is utilized to calculate the fidelities for \{4, 6, 8, 10\}-atom systems. 
The thermal dephasing (including the position uncertainty and the Doppler shifts) is considered during the numerical simulation by  repeatedly calculating fidelities based on different values of $R$ and $\delta^D$ (obeying the normal distribution), which induces the error bars (standard deviation) in Fig. \ref{Fidelity}.
For each simulated final density matrix $\rho_f$, we deduce fidelity $\mathcal{F}=\langle {\rm GHZ}_N| \rho_f | {\rm GHZ}_N\rangle=\frac{1}{2}(p_{A_N}+p_{\bar{A}_N}+c_N+c_N^*)$ \cite{lukinScience}, where $p_{A_N}(p_{\bar{A}_N})$ is  population on $|A_N\rangle=|d0d0\cdots d0\rangle$ $(|\bar{A}_N\rangle=|0d0d\cdots0d\rangle)$ and $c_N=\langle \bar{A}_N|\rho_f|A_N\rangle$ describes the coincidence of off-diagonal terms.
The fidelity of the 4-atom GHZ state is higher than $97\%$.
Even in the 10-atom system, the fidelity is still close to $85\%$.
Additionally, according to the prediction line in Fig. \ref{Fidelity}, the fidelity of 20-atom will be around 80\%.

\section{Scaling of $T_2$}\label{Scaling_T2}
\begin{figure}
\includegraphics[width=8cm]{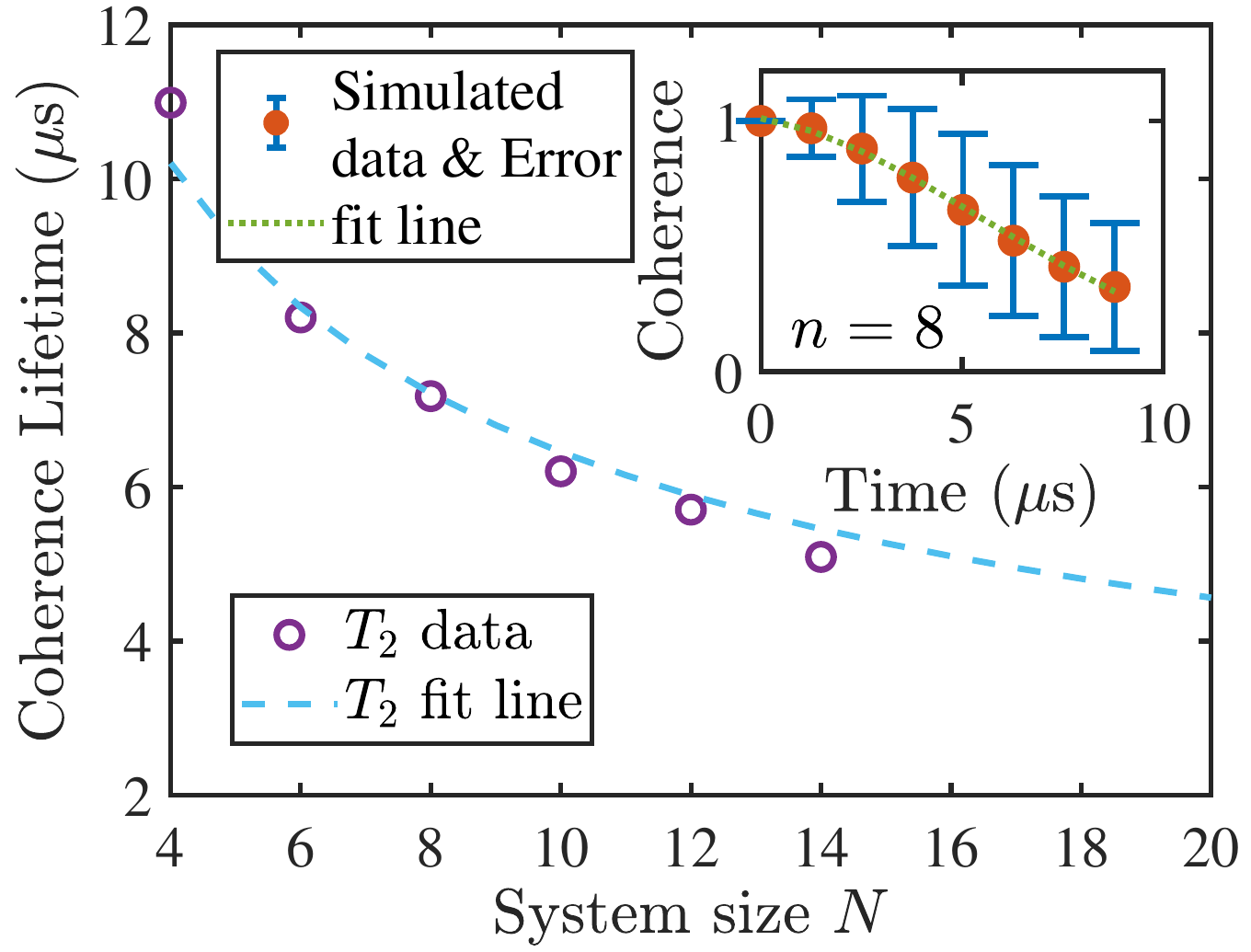}
\caption{Coherence lifetime ($T_2$) versus system size $N$.
The circles show the $T_2$ data inferred by Gaussian fittings of the coherence damping for systems with different sizes and the blue dashed line gives a predicted trend ($T_2\propto 1/\sqrt{N} $) for larger systems ($N\ge16$).
As an example, the inset exhibits the coherence damping of an $8$-atom system with a green dotted line obeying Gaussian fitting.
Similar to Fig. \ref{Fidelity}, the dots and error bars are mean values and standard deviations of coherence.}
\label{T2}
\end{figure}

To further benchmark the dephasing effect of present dressed-atom systems, we assume the initial state is $|{\rm GHZ}_N\rangle$ and turn off all the driving lases for a specific time $\tau$ and calculate the corresponding density matrix $\rho'(\tau)$.
By counting $2|\langle \bar{A}_N|\rho'(\tau)|A_N\rangle|$ as the coherence and defining the coherence lifetime $T_2$ satisfying $2|\langle \bar{A}_N|\rho'(T_2)|A_N\rangle|=e^{-1}$,
we show $T_2$ versus system size $N$ in Fig. \ref{T2}.
These $T_2$ data are around 3 times that of the observed data in the previous work \cite{lukinScience} without dressing atoms, at the same system size.

\begin{figure}
\includegraphics[width=7cm]{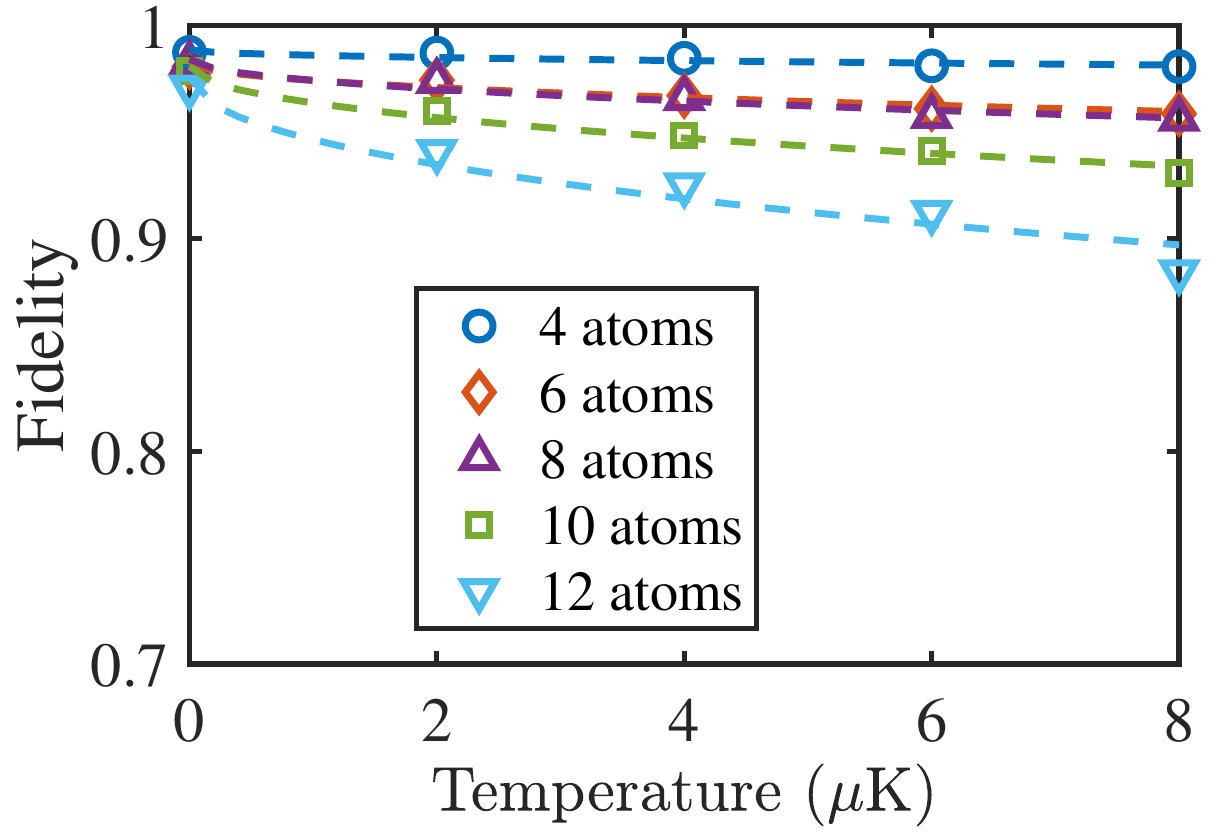}
\caption{Fidelities of GHZ states versus atomic temperature. 
The circles, diamonds, triangles, and squares represent the fidelity of \{4, 6, 8, 10, 12\}-atom GHZ states, respectively.
The dash lines correspondingly show the fitting results based on $\ln (\mathcal{F}-1/2) \propto -\sqrt{T}$ for different scales systems.
}
\label{FvsT}
\end{figure}

The above results come from the current cooling limited temperature $T=10$ $\mu$K of cold atoms \cite{lukinScience,lukin-256}. 
In the future, with the development of experimental techniques, the atoms may be cooled to lower temperatures. 
We here give an estimate of fidelity $\mathcal{F}$ versus atomic temperature $T$ to show the prospect of our scheme,
with the simulation results of \{4, 6, 8, 10\}-atom systems shown in Fig. \ref{FvsT}.
The relationship between fidelity $\mathcal{F}$ and decoherence time $T_2$ is $\ln(\mathcal{F}-1/2)\propto -1/T_2$.
The coherence $T_2$ results from the thermal motion described by the Boltzmann distribution and therefore satisfies $T_2 \propto 1/\sqrt{T}$.
The final relationship between fidelity $\mathcal{F}$ and atomic temperature $T$ is $\ln (\mathcal{F}-1/2) \propto -\sqrt{T}$, which coincides well with the simulation results in Fig. \ref{FvsT}, 
and also gives prediction that the fidelities of \{4, 6, 8, 10, 12\}-atom GHZ states can reach above 90\% when the atomic temperature is lower than 10 $\mu$K.

It is worth noting that the Rydberg state percentage $P_r$ of the dressed state is $25\%$.
According to the lifetime of $70S$ Rydberg state of $146$ $\mu$s \cite{Rblifetime,Dopplerdata-PRL}, the lifetime of the dressed state could extend to $573$ $\mu$s, 
long enough to ignore the depopulation effect factor as the preparation times of GHZ states around $10$ $\mu$s.
Additionally, we only consider the near-neighbor and next-neighbor interactions.

\section{Conclusion}\label{Conclusion}
We have explored the thermal-dephasing-tolerant generation of GHZ states in dressed-atom systems.
The thermal dephasing is suppressed here to 1/3 of the original performance in multi-body entangled systems \cite{lukinScience}, 
leading to solid fidelities of multi-body GHZ states.
As the dephasing factor becomes an important obstacle to large-scale quantum computation and quantum simulation based on neutral atoms, the present work may be referable to building dephasing-tolerant quantum tasks in large-scale systems.
Combining the Rydberg dressed state method with GPAPE optimal control techniques also provides an enlightening new perspective for robust handling of many-body Rydberg quantum simulations and quantum computation.

\section*{Acknowledgement}
This work was supported by the National Natural Science Foundation of China under Grants No. 11575045, No. 11874114, No. 11674060, No. 11805036, and the Natural Science Funds for Distinguished Young Scholar of Fujian Province under Grant No. 2020J06011, Project from Fuzhou University under Grant JG202001-2.
S. L. S. acknowledges support from National Natural Science Foundation of China under Grant No. 12274376 and major science and technology project of Henan Province under Grant No. 221100210400.
W. L. acknowledges support from the EPSRC through Grant No.~EP/W015641/1 and the British Council through an Industry Academia Collaborative Grant (No. IND/CONT/G/22-23/26).

\appendix
\section{Calculation dimension reduction through the Rydberg dressed blockade} \label{Appendix_A}
\begin{figure}
\includegraphics[width=8cm]{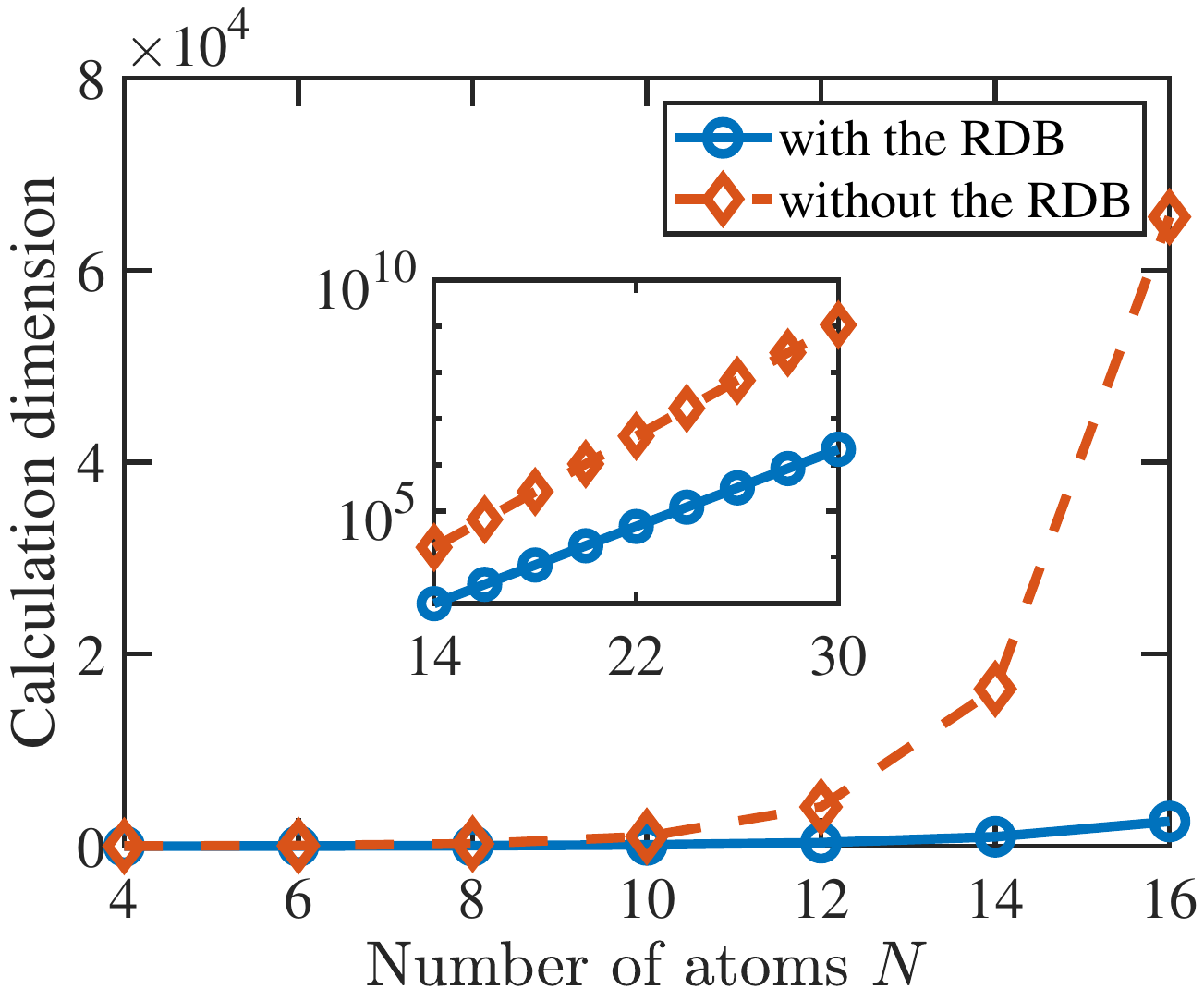}
\caption{Calculation dimension versus the number of atoms $N$ (a) with RDB and (b) without RDB effects. 
The inset shows the case of logarithmic coordinates when $N\in[14,30]$.}
\label{GBBN}
\end{figure}

In the main text, we use $H_{\rm mw}$ to  manipulate the dynamics of multiple atoms in subspace $\mathcal{S}_1=$\{$|0\rangle$, $|d\rangle$\}. 
Further, the subspace $\mathcal{S}_2=\{|dd\rangle_{i(i+1)}\}$ are forbidden because of the Rydberg dressed blockade (RDB) effect.
Therefore, the calculation space is locked to $\mathcal{S}_3=\complement_{\mathcal{S}_1}\mathcal{S}_2$, with
the calculation dimension reduced from $2^N$ to $\sum_{m=0}^{N/2}C_{N+1-m}^m$ (see Fig. \ref{GBBN}). 

\section{Adiabatic passage for the GHZ states generation}\label{Appendix_B}
\begin{figure*}
\includegraphics[width=13cm]{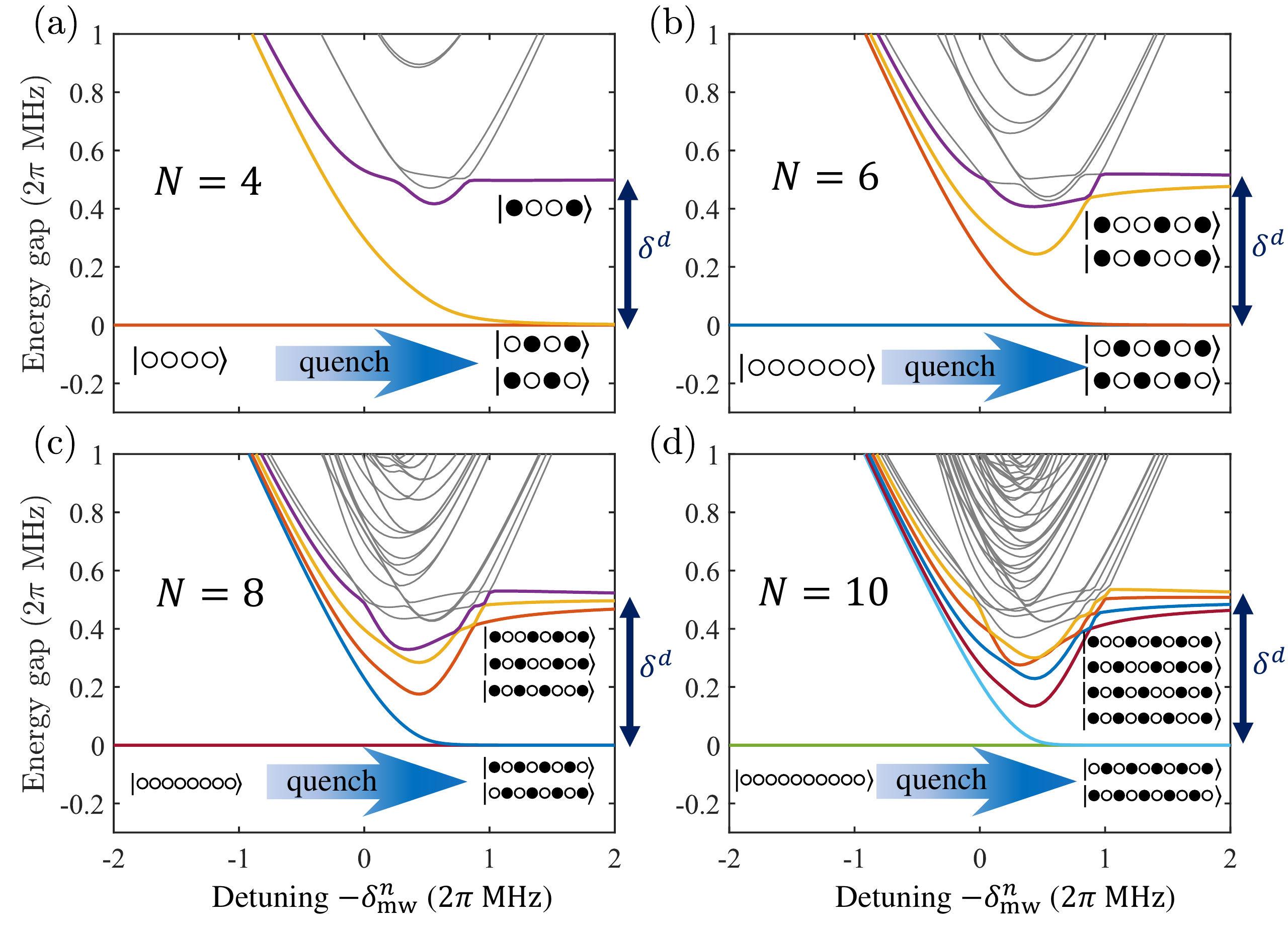}
\caption{ Energy gaps versus the microwave detuning $-\delta_{\rm mw}^n$ for the system size (a) $N=4$, (b) $N=6$, (c) $N=8$, and (d) $N=10$.
We choose $\Omega_{\rm mw}/(2\pi)=0.2$ MHz and $(\delta_{\rm mw}^{e}-\delta_{\rm mw}^{ne})/(2\pi)=\delta^d/(2\pi)=0.5$ MHz.
Formats $|\circ\rangle$ and $|\bullet\rangle$ represent $|0\rangle$ and $|d\rangle$ for visual intuition, respectively.
Some unimportant eigenstates are plotted as gray.
This graph refers to the presentation form of Fig. 1(b) in Ref. \cite{lukinScience}, with data differences.}
\label{adabatic_graph}
\end{figure*}

\begin{figure*}
\includegraphics[width=16cm]{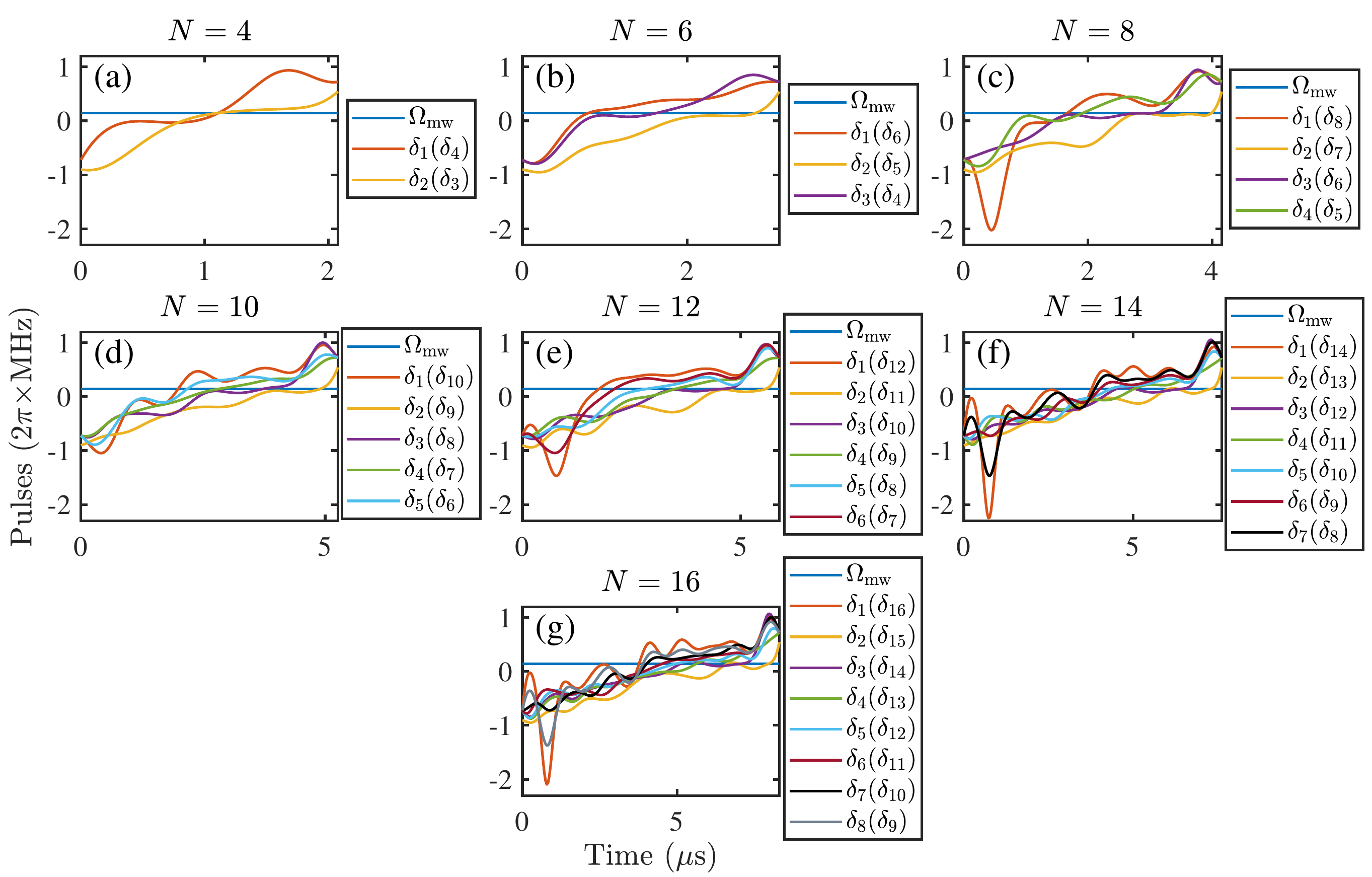}
\caption{(Color online) Optimized microwave pulses by the GRAPE in the time domain.(a-g) The microwave pulses for systems with different sizes ${N= \{4, 6, 8, 10, 12, 14, 16\}}$.}
\label{pulses}
\end{figure*}

\begin{figure*}
\includegraphics[width=16cm]{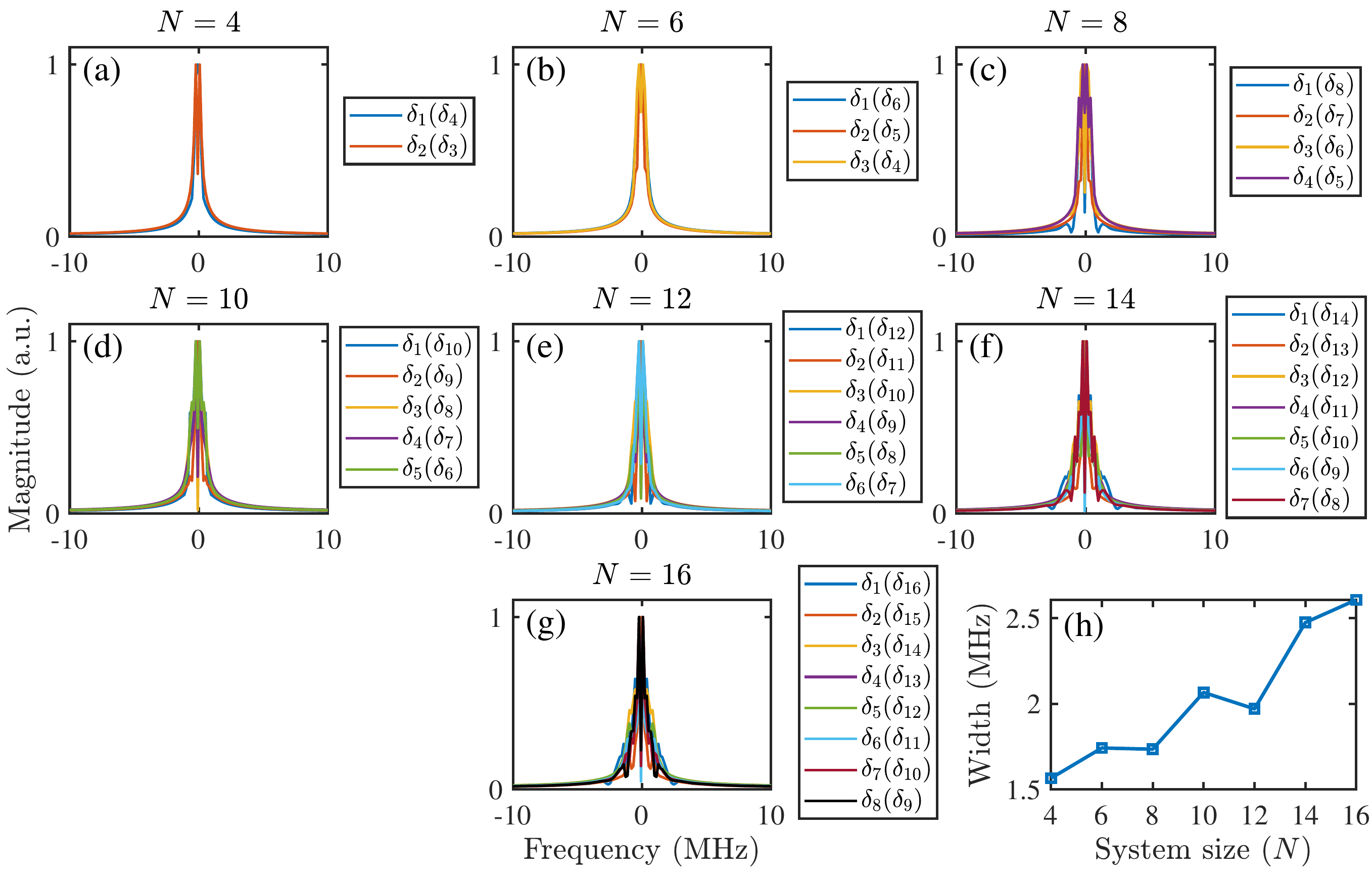}
\caption{(Color online) Optimized microwave pulses by the GRAPE in the frequency domain.
	(a-g) The microwave pulses for systems with different sizes ${N= \{4, 6, 8, 10, 12, 14, 16\}}$. 
	(h) Full width at 1/5 maximum versus the system size $ N$.}
\label{spectrum}
\end{figure*}

In the subspace $\mathcal{S}_2$ restricted by the RDB, there are at most $N/2$ atoms in dressed states.
When the detuning $-\delta_{\rm mw}^n$ are large negative values (compare to $\Omega_{\rm mw}$), the system ground state is  $|0000\cdots \rangle$.
When $-\delta_{\rm mw}^n$ are large positive values, the system ground state becomes degenerate with $N/2+1$ states with $N/2$ dressed-state excitation, for examples, ground states $\{|d00d\rangle,|0d0d\rangle,|d0d0\rangle\}$ when $N=4$
and $\{|d0d00d\rangle,|d00d0d\rangle,|0d0d0d\rangle,|d0d0d0\rangle\}$ when $N=6$.
Further, by differing the detuning $-\delta_{\rm mw}^n$ between the edge atoms and other atoms, one can separate such degenerate ground states with energy $\delta^d/(2\pi)=(\delta_{\rm mw}^{e}-\delta_{\rm mw}^{ne})/(2\pi)=0.5$ MHz.
Subsequently, the ground state when $-\delta_{\rm mw}^n$ are large positive values becomes only the GHZ state $| {\rm GHZ}_N\rangle$.
That is why we can adiabatically change the detuning $-\delta_{\rm mw}^n$ from large negative values to large positive values to prepare the GHZ states from the initial state $|0000\cdots \rangle$.
For visualization, we have drawn the diagram of adiabatic passages for systems with different sizes in Fig. \ref{adabatic_graph}.
Such an adiabatic method has been proved in the work of Ref. \cite{lukinScience}.

\section{GRAPE-optimized pulses for the GHZ states generation}\label{Appendix_C}
The optimized method we used is call  gradient ascent pulse engineering (GRAPE) \cite{GRAPE}.
We preset the control Hamiltonians as $H_h=|0\rangle_h\langle 0|+|0\rangle_{N+1-h}\langle 0|$ ($h=1,2,...,N/2$).
The corresponding control pulses are set as $f_h=\delta_{\rm mw}^{h}$.
The detailed derivation can be found in Ref. \cite{GRAPE}, and here we only give the specific optimization steps:

(1) Guess the initial microwave pulses $\Omega_{\rm mw}=0.14$ MHz, $f_{h\neq 1}=(10\Omega_{\rm mw}\cdot t/t_f-5\Omega_{\rm mw})$, and $f_{1}=f_{h\neq 1}-1.25\Omega_{\rm mw}$, 
where $t$ is the evolution time and $t_f$ is the final evolution time depended on the system size $N$.

(2) Calculate the evolution operator $U(t)$.

(3) Change new control pulses as
$f'_h=f_h-i \epsilon_h \cdot \Delta t \cdot {\rm Tr}\left\{[H_k,U(t)\rho_0U(t)^\dag]\cdot U(t)U(t_f)^\dag \rho_fU(t_f)U(t)^\dag \right \}$, with $\rho_f$ the desired state, i.e., the GHZ state.
Here $\epsilon_h=\Omega_{\rm mw}/2$  and $\Delta t=t_f/200$ are the correction strength and the time derivative, respectively. 

(4) Go to step (2) until the fidelity ${\rm Tr}[U(t)\rho_0U(t)^\dag \rho_f]$ reaches a certain requirement.

The optimization results are shown in Table I in the main text.
Note that the maximum absolute value of $\Omega_{\rm mw}$ does not exceed $2\pi \times 0.14$ MHz to ensure the stability of the RDB dynamics ($J\gg\Omega_{\rm mw}$). 

The optimized pulses in the time domain are shown in Fig. \ref{pulses}. Please note that the pulses look complicated because we place all the pulses into one subfigure. In fact, each atom has only two microwave pulses ${\Omega_{\rm mw}}$ and ${\delta_{\rm mw}^n}$. All of the addressing pulses in Fig. \ref{pulses} are well-realized experimentally \cite{lukinScience}.

As the system size $ N$ increases, small oscillations gradually appear in microwave pulses (see Fig. \ref{pulses}). To characterize these small amplitude oscillations, we plot the pulses in the frequency domain in Fig. \ref{spectrum}. We define frequency broadening as the full width at 1/5 maximum (similar to the full width at half maximum) to define the frequency width. Such widths are plotted in Fig. \ref{spectrum}(h), which gradually increases as the system size $ N$ increases. This reflects, to some extent, the difficulty of implementing such pulses experimentally because high irregular frequency microwaves usually have rising and falling edges problems. In any case, widths around 2 MHz frequency are entirely achievable using current experimental techniques \cite{lukinScience}.

\section{Selection of the preparation time of the GHZ states}\label{Appendix_D}

\begin{figure}
\includegraphics[width=8.6cm]{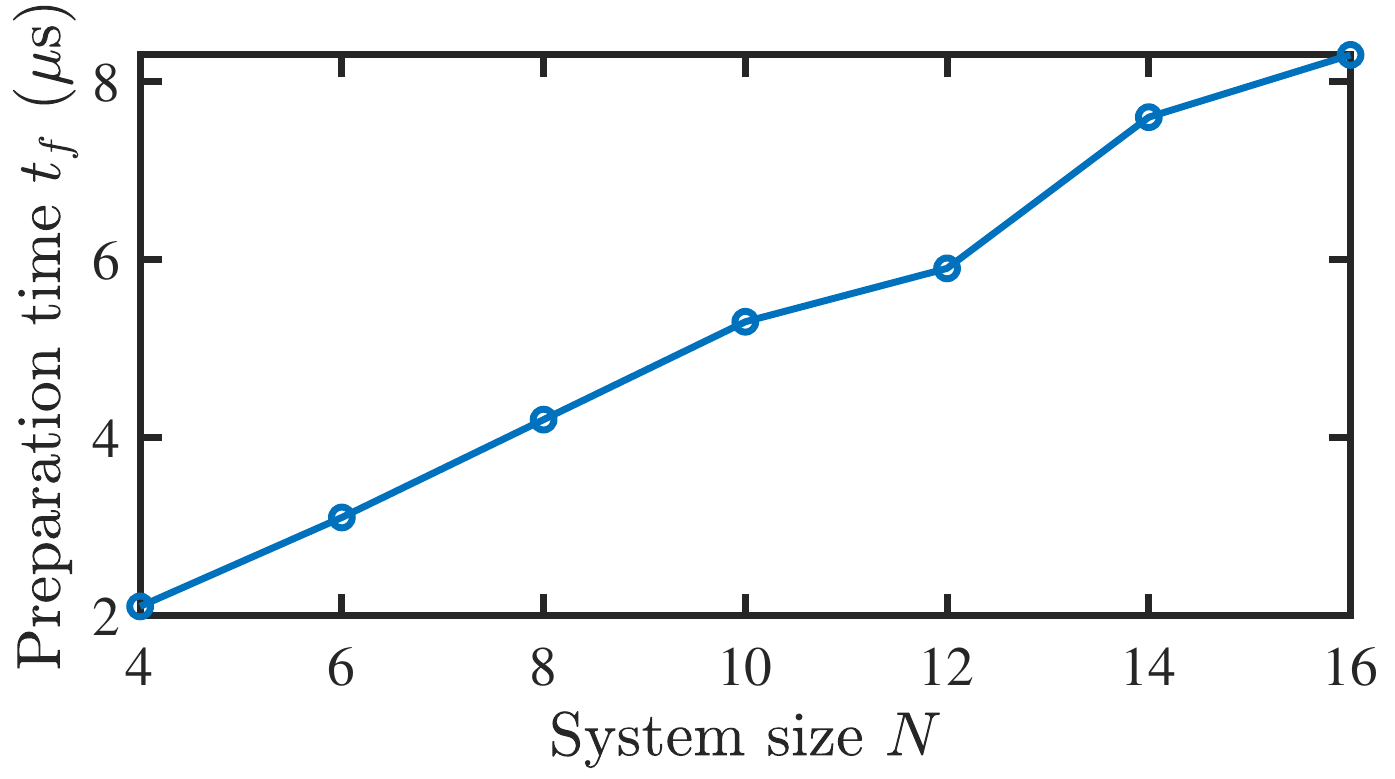}
\caption{Preparation time ${t_f}$ of the GHZ states generation versus System size $ N$. The curve has a trend with ${t_f=N/2}$ $ \mu$s.}
\label{tf_vs_N}
\end{figure}

Specifically, we note that the preparation time $t_f$ of the GHZ states linearly increases with system size (see Fig. \ref{tf_vs_N}). This is primarily because that the GRAPE method constantly corrects pulses to achieve higher fidelity during optimization. As the system size increases and the degree of freedom of the system becomes larger, it becomes more difficult to find a set of non-adiabatic pulses that achieve the fidelity threshold (0.99 in this work). Therefore, we appropriately increase the evolution time as the size of the system increases. A rough rule is that $t_f=N/2$ $\mu$s.

\section{Parameter choice of the simulation for Fig. 5 in the main text}\label{Appendix_E}
Figure \ref{FvsT} in  the main text demonstrates the relationship between fidelities of GHZ states $\cal F$ and atomic temperature $T$.
The selection of parameters for each temperature is different to ensure high fidelity $\cal F$.
Specifically, we choose $\{V,\Delta,|\Omega_{\rm mw}|_{\rm max}\}/(2\pi)=\{20,7.8,0.1\}$, $\{18,5.5,0.1\}$, $\{19,5.5,0.1\}$, $\{20,5.5,0.1\}$, and $\{20,5.5,0.1\}$ MHz for $T=0$, 2, 4, 6, and 8 $\mu$K, respectively.

\bibliography{catRDB_ref}

\begin{thebibliography}{58}%
\makeatletter
\providecommand \@ifxundefined [1]{%
 \@ifx{#1\undefined}
}%
\providecommand \@ifnum [1]{%
 \ifnum #1\expandafter \@firstoftwo
 \else \expandafter \@secondoftwo
 \fi
}%
\providecommand \@ifx [1]{%
 \ifx #1\expandafter \@firstoftwo
 \else \expandafter \@secondoftwo
 \fi
}%
\providecommand \natexlab [1]{#1}%
\providecommand \enquote  [1]{``#1''}%
\providecommand \bibnamefont  [1]{#1}%
\providecommand \bibfnamefont [1]{#1}%
\providecommand \citenamefont [1]{#1}%
\providecommand \href@noop [0]{\@secondoftwo}%
\providecommand \href [0]{\begingroup \@sanitize@url \@href}%
\providecommand \@href[1]{\@@startlink{#1}\@@href}%
\providecommand \@@href[1]{\endgroup#1\@@endlink}%
\providecommand \@sanitize@url [0]{\catcode `\\12\catcode `\$12\catcode
  `\&12\catcode `\#12\catcode `\^12\catcode `\_12\catcode `\%12\relax}%
\providecommand \@@startlink[1]{}%
\providecommand \@@endlink[0]{}%
\providecommand \url  [0]{\begingroup\@sanitize@url \@url }%
\providecommand \@url [1]{\endgroup\@href {#1}{\urlprefix }}%
\providecommand \urlprefix  [0]{URL }%
\providecommand \Eprint [0]{\href }%
\providecommand \doibase [0]{https://doi.org/}%
\providecommand \selectlanguage [0]{\@gobble}%
\providecommand \bibinfo  [0]{\@secondoftwo}%
\providecommand \bibfield  [0]{\@secondoftwo}%
\providecommand \translation [1]{[#1]}%
\providecommand \BibitemOpen [0]{}%
\providecommand \bibitemStop [0]{}%
\providecommand \bibitemNoStop [0]{.\EOS\space}%
\providecommand \EOS [0]{\spacefactor3000\relax}%
\providecommand \BibitemShut  [1]{\csname bibitem#1\endcsname}%
\let\auto@bib@innerbib\@empty
\bibitem [{\citenamefont {Saffman}\ \emph {et~al.}(2010)\citenamefont
  {Saffman}, \citenamefont {Walker},\ and\ \citenamefont
  {M\o{}lmer}}]{Saffman2010RMP}%
  \BibitemOpen
  \bibfield  {author} {\bibinfo {author} {\bibfnamefont {M.}~\bibnamefont
  {Saffman}}, \bibinfo {author} {\bibfnamefont {T.~G.}\ \bibnamefont
  {Walker}},\ and\ \bibinfo {author} {\bibfnamefont {K.}~\bibnamefont
  {M\o{}lmer}},\ }\bibfield  {title} {\bibinfo {title} {Quantum information
  with rydberg atoms},\ }\href {https://doi.org/10.1103/RevModPhys.82.2313}
  {\bibfield  {journal} {\bibinfo  {journal} {Rev. Mod. Phys.}\ }\textbf
  {\bibinfo {volume} {82}},\ \bibinfo {pages} {2313} (\bibinfo {year}
  {2010})}\BibitemShut {NoStop}%
\bibitem [{\citenamefont {Schauß}\ \emph {et~al.}(2012)\citenamefont
  {Schauß}, \citenamefont {Cheneau}, \citenamefont {Endres}, \citenamefont
  {Fukuhara}, \citenamefont {Hild}, \citenamefont {Omran}, \citenamefont
  {Pohl}, \citenamefont {Gross}, \citenamefont {Kuhr},\ and\ \citenamefont
  {Bloch}}]{Schaus2012Nature}%
  \BibitemOpen
  \bibfield  {author} {\bibinfo {author} {\bibfnamefont {P.}~\bibnamefont
  {Schauß}}, \bibinfo {author} {\bibfnamefont {M.}~\bibnamefont {Cheneau}},
  \bibinfo {author} {\bibfnamefont {M.}~\bibnamefont {Endres}}, \bibinfo
  {author} {\bibfnamefont {T.}~\bibnamefont {Fukuhara}}, \bibinfo {author}
  {\bibfnamefont {S.}~\bibnamefont {Hild}}, \bibinfo {author} {\bibfnamefont
  {A.}~\bibnamefont {Omran}}, \bibinfo {author} {\bibfnamefont
  {T.}~\bibnamefont {Pohl}}, \bibinfo {author} {\bibfnamefont {C.}~\bibnamefont
  {Gross}}, \bibinfo {author} {\bibfnamefont {S.}~\bibnamefont {Kuhr}},\ and\
  \bibinfo {author} {\bibfnamefont {I.}~\bibnamefont {Bloch}},\ }\bibfield
  {title} {\bibinfo {title} {Observation of spatially ordered structures in a
  two-dimensional {Rydberg} gas},\ }\href {https://doi.org/10.1038/nature11596}
  {\bibfield  {journal} {\bibinfo  {journal} {Nature}\ }\textbf {\bibinfo
  {volume} {491}},\ \bibinfo {pages} {87} (\bibinfo {year} {2012})}\BibitemShut
  {NoStop}%
\bibitem [{\citenamefont {Hofmann}\ \emph {et~al.}(2013)\citenamefont
  {Hofmann}, \citenamefont {G\"unter}, \citenamefont {Schempp}, \citenamefont
  {Robert-de Saint-Vincent}, \citenamefont {G\"arttner}, \citenamefont {Evers},
  \citenamefont {Whitlock},\ and\ \citenamefont
  {Weidem\"uller}}]{Hofmann2013PRL}%
  \BibitemOpen
  \bibfield  {author} {\bibinfo {author} {\bibfnamefont {C.~S.}\ \bibnamefont
  {Hofmann}}, \bibinfo {author} {\bibfnamefont {G.}~\bibnamefont {G\"unter}},
  \bibinfo {author} {\bibfnamefont {H.}~\bibnamefont {Schempp}}, \bibinfo
  {author} {\bibfnamefont {M.}~\bibnamefont {Robert-de Saint-Vincent}},
  \bibinfo {author} {\bibfnamefont {M.}~\bibnamefont {G\"arttner}}, \bibinfo
  {author} {\bibfnamefont {J.}~\bibnamefont {Evers}}, \bibinfo {author}
  {\bibfnamefont {S.}~\bibnamefont {Whitlock}},\ and\ \bibinfo {author}
  {\bibfnamefont {M.}~\bibnamefont {Weidem\"uller}},\ }\bibfield  {title}
  {\bibinfo {title} {Sub-poissonian statistics of {Rydberg}-interacting
  dark-state polaritons},\ }\href
  {https://doi.org/10.1103/PhysRevLett.110.203601} {\bibfield  {journal}
  {\bibinfo  {journal} {Phys. Rev. Lett.}\ }\textbf {\bibinfo {volume} {110}},\
  \bibinfo {pages} {203601} (\bibinfo {year} {2013})}\BibitemShut {NoStop}%
\bibitem [{\citenamefont {Ebadi}\ \emph {et~al.}(2021)\citenamefont {Ebadi},
  \citenamefont {Wang}, \citenamefont {Levine}, \citenamefont {Keesling},
  \citenamefont {Semeghini}, \citenamefont {Omran}, \citenamefont {Bluvstein},
  \citenamefont {Samajdar}, \citenamefont {Pichler}, \citenamefont {Ho},
  \citenamefont {Choi}, \citenamefont {Sachdev}, \citenamefont {Greiner},
  \citenamefont {Vuleti\'{c}},\ and\ \citenamefont {Lukin}}]{lukin-256}%
  \BibitemOpen
  \bibfield  {author} {\bibinfo {author} {\bibfnamefont {S.}~\bibnamefont
  {Ebadi}}, \bibinfo {author} {\bibfnamefont {T.~T.}\ \bibnamefont {Wang}},
  \bibinfo {author} {\bibfnamefont {H.}~\bibnamefont {Levine}}, \bibinfo
  {author} {\bibfnamefont {A.}~\bibnamefont {Keesling}}, \bibinfo {author}
  {\bibfnamefont {G.}~\bibnamefont {Semeghini}}, \bibinfo {author}
  {\bibfnamefont {A.}~\bibnamefont {Omran}}, \bibinfo {author} {\bibfnamefont
  {D.}~\bibnamefont {Bluvstein}}, \bibinfo {author} {\bibfnamefont
  {R.}~\bibnamefont {Samajdar}}, \bibinfo {author} {\bibfnamefont
  {H.}~\bibnamefont {Pichler}}, \bibinfo {author} {\bibfnamefont {W.~W.}\
  \bibnamefont {Ho}}, \bibinfo {author} {\bibfnamefont {S.}~\bibnamefont
  {Choi}}, \bibinfo {author} {\bibfnamefont {S.}~\bibnamefont {Sachdev}},
  \bibinfo {author} {\bibfnamefont {M.}~\bibnamefont {Greiner}}, \bibinfo
  {author} {\bibfnamefont {V.}~\bibnamefont {Vuleti\'{c}}},\ and\ \bibinfo
  {author} {\bibfnamefont {M.~D.}\ \bibnamefont {Lukin}},\ }\bibfield  {title}
  {\bibinfo {title} {{Quantum phases of matter on a 256-atom programmable
  quantum simulator}},\ }\href {https://doi.org/10.1038/s41586-021-03582-4}
  {\bibfield  {journal} {\bibinfo  {journal} {Nature}\ }\textbf {\bibinfo
  {volume} {595}},\ \bibinfo {pages} {227} (\bibinfo {year}
  {2021})}\BibitemShut {NoStop}%
\bibitem [{\citenamefont {Bluvstein}\ \emph {et~al.}(2022)\citenamefont
  {Bluvstein}, \citenamefont {Levine}, \citenamefont {Semeghini}, \citenamefont
  {Wang}, \citenamefont {Ebadi}, \citenamefont {Kalinowski}, \citenamefont
  {Keesling}, \citenamefont {Maskara}, \citenamefont {Pichler}, \citenamefont
  {Greiner}, \citenamefont {Vuleti\'c},\ and\ \citenamefont
  {Lukin}}]{Bluvstein2022Nature}%
  \BibitemOpen
  \bibfield  {author} {\bibinfo {author} {\bibfnamefont {D.}~\bibnamefont
  {Bluvstein}}, \bibinfo {author} {\bibfnamefont {H.}~\bibnamefont {Levine}},
  \bibinfo {author} {\bibfnamefont {G.}~\bibnamefont {Semeghini}}, \bibinfo
  {author} {\bibfnamefont {T.~T.}\ \bibnamefont {Wang}}, \bibinfo {author}
  {\bibfnamefont {S.}~\bibnamefont {Ebadi}}, \bibinfo {author} {\bibfnamefont
  {M.}~\bibnamefont {Kalinowski}}, \bibinfo {author} {\bibfnamefont
  {A.}~\bibnamefont {Keesling}}, \bibinfo {author} {\bibfnamefont
  {N.}~\bibnamefont {Maskara}}, \bibinfo {author} {\bibfnamefont
  {H.}~\bibnamefont {Pichler}}, \bibinfo {author} {\bibfnamefont
  {M.}~\bibnamefont {Greiner}}, \bibinfo {author} {\bibfnamefont
  {V.}~\bibnamefont {Vuleti\'c}},\ and\ \bibinfo {author} {\bibfnamefont
  {M.~D.}\ \bibnamefont {Lukin}},\ }\bibfield  {title} {\bibinfo {title} {A
  quantum processor based on coherent transport of entangled atom arrays},\
  }\href {https://doi.org/10.1038/s41586-022-04592-6} {\bibfield  {journal}
  {\bibinfo  {journal} {Nature}\ }\textbf {\bibinfo {volume} {604}},\ \bibinfo
  {pages} {451} (\bibinfo {year} {2022})}\BibitemShut {NoStop}%
\bibitem [{\citenamefont {Henkel}\ \emph {et~al.}(2010)\citenamefont {Henkel},
  \citenamefont {Nath},\ and\ \citenamefont {Pohl}}]{Henkel2010PRL}%
  \BibitemOpen
  \bibfield  {author} {\bibinfo {author} {\bibfnamefont {N.}~\bibnamefont
  {Henkel}}, \bibinfo {author} {\bibfnamefont {R.}~\bibnamefont {Nath}},\ and\
  \bibinfo {author} {\bibfnamefont {T.}~\bibnamefont {Pohl}},\ }\bibfield
  {title} {\bibinfo {title} {Three-dimensional roton excitations and supersolid
  formation in rydberg-excited bose-einstein condensates},\ }\href
  {https://doi.org/10.1103/PhysRevLett.104.195302} {\bibfield  {journal}
  {\bibinfo  {journal} {Phys. Rev. Lett.}\ }\textbf {\bibinfo {volume} {104}},\
  \bibinfo {pages} {195302} (\bibinfo {year} {2010})}\BibitemShut {NoStop}%
\bibitem [{\citenamefont {Pupillo}\ \emph {et~al.}(2010)\citenamefont
  {Pupillo}, \citenamefont {Micheli}, \citenamefont {Boninsegni}, \citenamefont
  {Lesanovsky},\ and\ \citenamefont {Zoller}}]{dressed-PRL}%
  \BibitemOpen
  \bibfield  {author} {\bibinfo {author} {\bibfnamefont {G.}~\bibnamefont
  {Pupillo}}, \bibinfo {author} {\bibfnamefont {A.}~\bibnamefont {Micheli}},
  \bibinfo {author} {\bibfnamefont {M.}~\bibnamefont {Boninsegni}}, \bibinfo
  {author} {\bibfnamefont {I.}~\bibnamefont {Lesanovsky}},\ and\ \bibinfo
  {author} {\bibfnamefont {P.}~\bibnamefont {Zoller}},\ }\bibfield  {title}
  {\bibinfo {title} {Strongly correlated gases of {Rydberg}-dressed atoms:
  Quantum and classical dynamics},\ }\href
  {https://doi.org/10.1103/PhysRevLett.104.223002} {\bibfield  {journal}
  {\bibinfo  {journal} {Phys. Rev. Lett.}\ }\textbf {\bibinfo {volume} {104}},\
  \bibinfo {pages} {223002} (\bibinfo {year} {2010})}\BibitemShut {NoStop}%
\bibitem [{\citenamefont {Bouchoule}\ and\ \citenamefont
  {M\o{}lmer}(2002)}]{Molmer-dressedGB-PRA}%
  \BibitemOpen
  \bibfield  {author} {\bibinfo {author} {\bibfnamefont {I.}~\bibnamefont
  {Bouchoule}}\ and\ \bibinfo {author} {\bibfnamefont {K.}~\bibnamefont
  {M\o{}lmer}},\ }\bibfield  {title} {\bibinfo {title} {Spin squeezing of atoms
  by the dipole interaction in virtually excited {Rydberg} states},\ }\href
  {https://doi.org/10.1103/PhysRevA.65.041803} {\bibfield  {journal} {\bibinfo
  {journal} {Phys. Rev. A}\ }\textbf {\bibinfo {volume} {65}},\ \bibinfo
  {pages} {041803(R)} (\bibinfo {year} {2002})}\BibitemShut {NoStop}%
\bibitem [{\citenamefont {Johnson}\ and\ \citenamefont
  {Rolston}(2010)}]{Johnson-dressedGB-PRA}%
  \BibitemOpen
  \bibfield  {author} {\bibinfo {author} {\bibfnamefont {J.~E.}\ \bibnamefont
  {Johnson}}\ and\ \bibinfo {author} {\bibfnamefont {S.~L.}\ \bibnamefont
  {Rolston}},\ }\bibfield  {title} {\bibinfo {title} {Interactions between
  {Rydberg}-dressed atoms},\ }\href
  {https://doi.org/10.1103/PhysRevA.82.033412} {\bibfield  {journal} {\bibinfo
  {journal} {Phys. Rev. A}\ }\textbf {\bibinfo {volume} {82}},\ \bibinfo
  {pages} {033412} (\bibinfo {year} {2010})}\BibitemShut {NoStop}%
\bibitem [{\citenamefont {Balewski}\ \emph {et~al.}(2014)\citenamefont
  {Balewski}, \citenamefont {Krupp}, \citenamefont {Gaj}, \citenamefont
  {Hofferberth}, \citenamefont {L\"{o}w},\ and\ \citenamefont
  {Pfau}}]{Balewski-dressedGB-NJP}%
  \BibitemOpen
  \bibfield  {author} {\bibinfo {author} {\bibfnamefont {J.~B.}\ \bibnamefont
  {Balewski}}, \bibinfo {author} {\bibfnamefont {A.~T.}\ \bibnamefont {Krupp}},
  \bibinfo {author} {\bibfnamefont {A.}~\bibnamefont {Gaj}}, \bibinfo {author}
  {\bibfnamefont {S.}~\bibnamefont {Hofferberth}}, \bibinfo {author}
  {\bibfnamefont {R.}~\bibnamefont {L\"{o}w}},\ and\ \bibinfo {author}
  {\bibfnamefont {T.}~\bibnamefont {Pfau}},\ }\bibfield  {title} {\bibinfo
  {title} {{Rydberg} dressing: understanding of collective many-body effects
  and implications for experiments},\ }\href
  {https://doi.org/10.1088/1367-2630/16/6/063012} {\bibfield  {journal}
  {\bibinfo  {journal} {New J. Phys.}\ }\textbf {\bibinfo {volume} {16}},\
  \bibinfo {pages} {063012} (\bibinfo {year} {2014})}\BibitemShut {NoStop}%
\bibitem [{\citenamefont {Yan}\ \emph {et~al.}(2011)\citenamefont {Yan},
  \citenamefont {Huse},\ and\ \citenamefont {White}}]{dressed-Science}%
  \BibitemOpen
  \bibfield  {author} {\bibinfo {author} {\bibfnamefont {S.}~\bibnamefont
  {Yan}}, \bibinfo {author} {\bibfnamefont {D.~A.}\ \bibnamefont {Huse}},\ and\
  \bibinfo {author} {\bibfnamefont {S.~R.}\ \bibnamefont {White}},\ }\bibfield
  {title} {\bibinfo {title} {Spin-liquid ground state of the ${S} = 1/2$ kagome
  {Heisenberg} antiferromagnet},\ }\href
  {https://doi.org/10.1126/science.1201080} {\bibfield  {journal} {\bibinfo
  {journal} {Science}\ }\textbf {\bibinfo {volume} {332}},\ \bibinfo {pages}
  {1173} (\bibinfo {year} {2011})}\BibitemShut {NoStop}%
\bibitem [{\citenamefont {Gil}\ \emph {et~al.}(2014)\citenamefont {Gil},
  \citenamefont {Mukherjee}, \citenamefont {Bridge}, \citenamefont {Jones},\
  and\ \citenamefont {Pohl}}]{dressedspin}%
  \BibitemOpen
  \bibfield  {author} {\bibinfo {author} {\bibfnamefont {L.~I.~R.}\
  \bibnamefont {Gil}}, \bibinfo {author} {\bibfnamefont {R.}~\bibnamefont
  {Mukherjee}}, \bibinfo {author} {\bibfnamefont {E.~M.}\ \bibnamefont
  {Bridge}}, \bibinfo {author} {\bibfnamefont {M.~P.~A.}\ \bibnamefont
  {Jones}},\ and\ \bibinfo {author} {\bibfnamefont {T.}~\bibnamefont {Pohl}},\
  }\bibfield  {title} {\bibinfo {title} {{Spin Squeezing in a Rydberg Lattice
  Clock}},\ }\href {https://doi.org/10.1103/PhysRevLett.112.103601} {\bibfield
  {journal} {\bibinfo  {journal} {Phys. Rev. Lett.}\ }\textbf {\bibinfo
  {volume} {112}},\ \bibinfo {pages} {103601} (\bibinfo {year}
  {2014})}\BibitemShut {NoStop}%
\bibitem [{\citenamefont {Keating}\ \emph {et~al.}(2013)\citenamefont
  {Keating}, \citenamefont {Goyal}, \citenamefont {Jau}, \citenamefont
  {Biedermann}, \citenamefont {Landahl},\ and\ \citenamefont
  {Deutsch}}]{dressedQC1}%
  \BibitemOpen
  \bibfield  {author} {\bibinfo {author} {\bibfnamefont {T.}~\bibnamefont
  {Keating}}, \bibinfo {author} {\bibfnamefont {K.}~\bibnamefont {Goyal}},
  \bibinfo {author} {\bibfnamefont {Y.-Y.}\ \bibnamefont {Jau}}, \bibinfo
  {author} {\bibfnamefont {G.~W.}\ \bibnamefont {Biedermann}}, \bibinfo
  {author} {\bibfnamefont {A.~J.}\ \bibnamefont {Landahl}},\ and\ \bibinfo
  {author} {\bibfnamefont {I.~H.}\ \bibnamefont {Deutsch}},\ }\bibfield
  {title} {\bibinfo {title} {{Adiabatic quantum computation with
  Rydberg-dressed atoms}},\ }\href {https://doi.org/10.1103/PhysRevA.87.052314}
  {\bibfield  {journal} {\bibinfo  {journal} {Phys. Rev. A}\ }\textbf {\bibinfo
  {volume} {87}},\ \bibinfo {pages} {052314} (\bibinfo {year}
  {2013})}\BibitemShut {NoStop}%
\bibitem [{\citenamefont {Keating}\ \emph {et~al.}(2015)\citenamefont
  {Keating}, \citenamefont {Cook}, \citenamefont {Hankin}, \citenamefont {Jau},
  \citenamefont {Biedermann},\ and\ \citenamefont {Deutsch}}]{dressedQC2}%
  \BibitemOpen
  \bibfield  {author} {\bibinfo {author} {\bibfnamefont {T.}~\bibnamefont
  {Keating}}, \bibinfo {author} {\bibfnamefont {R.~L.}\ \bibnamefont {Cook}},
  \bibinfo {author} {\bibfnamefont {A.~M.}\ \bibnamefont {Hankin}}, \bibinfo
  {author} {\bibfnamefont {Y.-Y.}\ \bibnamefont {Jau}}, \bibinfo {author}
  {\bibfnamefont {G.~W.}\ \bibnamefont {Biedermann}},\ and\ \bibinfo {author}
  {\bibfnamefont {I.~H.}\ \bibnamefont {Deutsch}},\ }\bibfield  {title}
  {\bibinfo {title} {{Robust quantum logic in neutral atoms via adiabatic
  Rydberg dressing}},\ }\href {https://doi.org/10.1103/PhysRevA.91.012337}
  {\bibfield  {journal} {\bibinfo  {journal} {Phys. Rev. A}\ }\textbf {\bibinfo
  {volume} {91}},\ \bibinfo {pages} {012337} (\bibinfo {year}
  {2015})}\BibitemShut {NoStop}%
\bibitem [{\citenamefont {Khazali}\ \emph {et~al.}(2016)\citenamefont
  {Khazali}, \citenamefont {Lau}, \citenamefont {Humeniuk},\ and\ \citenamefont
  {Simon}}]{dress_cat_1}%
  \BibitemOpen
  \bibfield  {author} {\bibinfo {author} {\bibfnamefont {M.}~\bibnamefont
  {Khazali}}, \bibinfo {author} {\bibfnamefont {H.~W.}\ \bibnamefont {Lau}},
  \bibinfo {author} {\bibfnamefont {A.}~\bibnamefont {Humeniuk}},\ and\
  \bibinfo {author} {\bibfnamefont {C.}~\bibnamefont {Simon}},\ }\bibfield
  {title} {\bibinfo {title} {Large energy superpositions via {Rydberg}
  dressing},\ }\href {https://doi.org/10.1103/PhysRevA.94.023408} {\bibfield
  {journal} {\bibinfo  {journal} {Phys. Rev. A}\ }\textbf {\bibinfo {volume}
  {94}},\ \bibinfo {pages} {023408} (\bibinfo {year} {2016})}\BibitemShut
  {NoStop}%
\bibitem [{\citenamefont {Khazali}(2018)}]{dress_cat_2}%
  \BibitemOpen
  \bibfield  {author} {\bibinfo {author} {\bibfnamefont {M.}~\bibnamefont
  {Khazali}},\ }\bibfield  {title} {\bibinfo {title} {Progress towards
  macroscopic spin and mechanical superposition via {Rydberg} interaction},\
  }\href {https://doi.org/10.1103/PhysRevA.98.043836} {\bibfield  {journal}
  {\bibinfo  {journal} {Phys. Rev. A}\ }\textbf {\bibinfo {volume} {98}},\
  \bibinfo {pages} {043836} (\bibinfo {year} {2018})}\BibitemShut {NoStop}%
\bibitem [{\citenamefont {Jau}\ \emph {et~al.}(2015)\citenamefont {Jau},
  \citenamefont {Hankin}, \citenamefont {Keating}, \citenamefont {Deutsch},\
  and\ \citenamefont {Biedermann}}]{Jau-GB-NP}%
  \BibitemOpen
  \bibfield  {author} {\bibinfo {author} {\bibfnamefont {Y.-Y.}\ \bibnamefont
  {Jau}}, \bibinfo {author} {\bibfnamefont {A.~M.}\ \bibnamefont {Hankin}},
  \bibinfo {author} {\bibfnamefont {T.}~\bibnamefont {Keating}}, \bibinfo
  {author} {\bibfnamefont {I.~H.}\ \bibnamefont {Deutsch}},\ and\ \bibinfo
  {author} {\bibfnamefont {G.~W.}\ \bibnamefont {Biedermann}},\ }\bibfield
  {title} {\bibinfo {title} {Entangling atomic spins with a {Rydberg}-dressed
  spin-flip blockade},\ }\href {https://doi.org/10.1038/nphys3487} {\bibfield
  {journal} {\bibinfo  {journal} {Nat. Phys.}\ }\textbf {\bibinfo {volume}
  {12}},\ \bibinfo {pages} {71} (\bibinfo {year} {2015})}\BibitemShut {NoStop}%
\bibitem [{\citenamefont {Shao}\ \emph {et~al.}(2017)\citenamefont {Shao},
  \citenamefont {Li}, \citenamefont {Ji}, \citenamefont {Wu},\ and\
  \citenamefont {Yi}}]{shao-GB}%
  \BibitemOpen
  \bibfield  {author} {\bibinfo {author} {\bibfnamefont {X.~Q.}\ \bibnamefont
  {Shao}}, \bibinfo {author} {\bibfnamefont {D.~X.}\ \bibnamefont {Li}},
  \bibinfo {author} {\bibfnamefont {Y.~Q.}\ \bibnamefont {Ji}}, \bibinfo
  {author} {\bibfnamefont {J.~H.}\ \bibnamefont {Wu}},\ and\ \bibinfo {author}
  {\bibfnamefont {X.~X.}\ \bibnamefont {Yi}},\ }\bibfield  {title} {\bibinfo
  {title} {Ground-state blockade of {Rydberg} atoms and application in
  entanglement generation},\ }\href
  {https://doi.org/10.1103/PhysRevA.96.012328} {\bibfield  {journal} {\bibinfo
  {journal} {Phys. Rev. A}\ }\textbf {\bibinfo {volume} {96}},\ \bibinfo
  {pages} {012328} (\bibinfo {year} {2017})}\BibitemShut {NoStop}%
\bibitem [{\citenamefont {Li}\ \emph {et~al.}(2019)\citenamefont {Li},
  \citenamefont {Zheng},\ and\ \citenamefont {Shao}}]{shao-GB-GHZ3}%
  \BibitemOpen
  \bibfield  {author} {\bibinfo {author} {\bibfnamefont {D.-X.}\ \bibnamefont
  {Li}}, \bibinfo {author} {\bibfnamefont {T.-Y.}\ \bibnamefont {Zheng}},\ and\
  \bibinfo {author} {\bibfnamefont {X.-Q.}\ \bibnamefont {Shao}},\ }\bibfield
  {title} {\bibinfo {title} {Adiabatic preparation of multipartite {GHZ} states
  via {Rydberg} ground-state blockade},\ }\href
  {https://doi.org/10.1364/oe.27.020874} {\bibfield  {journal} {\bibinfo
  {journal} {Opt. Express}\ }\textbf {\bibinfo {volume} {27}},\ \bibinfo
  {pages} {20874} (\bibinfo {year} {2019})}\BibitemShut {NoStop}%
\bibitem [{\citenamefont {Greenberger}\ \emph {et~al.}(1990)\citenamefont
  {Greenberger}, \citenamefont {Horne}, \citenamefont {Shimony},\ and\
  \citenamefont {Zeilinger}}]{GHZ}%
  \BibitemOpen
  \bibfield  {author} {\bibinfo {author} {\bibfnamefont {D.~M.}\ \bibnamefont
  {Greenberger}}, \bibinfo {author} {\bibfnamefont {M.~A.}\ \bibnamefont
  {Horne}}, \bibinfo {author} {\bibfnamefont {A.}~\bibnamefont {Shimony}},\
  and\ \bibinfo {author} {\bibfnamefont {A.}~\bibnamefont {Zeilinger}},\
  }\bibfield  {title} {\bibinfo {title} {Bell’s theorem without
  inequalities},\ }\href {https://doi.org/10.1119/1.16243} {\bibfield
  {journal} {\bibinfo  {journal} {Am. J Phys.}\ }\textbf {\bibinfo {volume}
  {58}},\ \bibinfo {pages} {1131} (\bibinfo {year} {1990})}\BibitemShut
  {NoStop}%
\bibitem [{\citenamefont {Agarwal}\ \emph {et~al.}(1997)\citenamefont
  {Agarwal}, \citenamefont {Puri},\ and\ \citenamefont {Singh}}]{atomcat}%
  \BibitemOpen
  \bibfield  {author} {\bibinfo {author} {\bibfnamefont {G.~S.}\ \bibnamefont
  {Agarwal}}, \bibinfo {author} {\bibfnamefont {R.~R.}\ \bibnamefont {Puri}},\
  and\ \bibinfo {author} {\bibfnamefont {R.~P.}\ \bibnamefont {Singh}},\
  }\bibfield  {title} {\bibinfo {title} {{Atomic Schr\"odinger cat states}},\
  }\href {https://doi.org/10.1103/PhysRevA.56.2249} {\bibfield  {journal}
  {\bibinfo  {journal} {Phys. Rev. A}\ }\textbf {\bibinfo {volume} {56}},\
  \bibinfo {pages} {2249} (\bibinfo {year} {1997})}\BibitemShut {NoStop}%
\bibitem [{\citenamefont {Gerry}\ and\ \citenamefont {Grobe}(1997)}]{atomcat2}%
  \BibitemOpen
  \bibfield  {author} {\bibinfo {author} {\bibfnamefont {C.~C.}\ \bibnamefont
  {Gerry}}\ and\ \bibinfo {author} {\bibfnamefont {R.}~\bibnamefont {Grobe}},\
  }\bibfield  {title} {\bibinfo {title} {{Generation and properties of
  collective atomic Schr\"odinger-cat states}},\ }\href
  {https://doi.org/10.1103/PhysRevA.56.2390} {\bibfield  {journal} {\bibinfo
  {journal} {Phys. Rev. A}\ }\textbf {\bibinfo {volume} {56}},\ \bibinfo
  {pages} {2390} (\bibinfo {year} {1997})}\BibitemShut {NoStop}%
\bibitem [{\citenamefont {Chen}\ \emph {et~al.}(2021)\citenamefont {Chen},
  \citenamefont {Qin}, \citenamefont {Wang}, \citenamefont {Miranowicz},\ and\
  \citenamefont {Nori}}]{cat_interest_1}%
  \BibitemOpen
  \bibfield  {author} {\bibinfo {author} {\bibfnamefont {Y.-H.}\ \bibnamefont
  {Chen}}, \bibinfo {author} {\bibfnamefont {W.}~\bibnamefont {Qin}}, \bibinfo
  {author} {\bibfnamefont {X.}~\bibnamefont {Wang}}, \bibinfo {author}
  {\bibfnamefont {A.}~\bibnamefont {Miranowicz}},\ and\ \bibinfo {author}
  {\bibfnamefont {F.}~\bibnamefont {Nori}},\ }\bibfield  {title} {\bibinfo
  {title} {Shortcuts to adiabaticity for the quantum {Rabi} model: Efficient
  generation of giant entangled cat states via parametric amplification},\
  }\href {https://doi.org/10.1103/PhysRevLett.126.023602} {\bibfield  {journal}
  {\bibinfo  {journal} {Phys. Rev. Lett.}\ }\textbf {\bibinfo {volume} {126}},\
  \bibinfo {pages} {023602} (\bibinfo {year} {2021})}\BibitemShut {NoStop}%
\bibitem [{\citenamefont {Qin}\ \emph {et~al.}(2021)\citenamefont {Qin},
  \citenamefont {Miranowicz}, \citenamefont {Jing},\ and\ \citenamefont
  {Nori}}]{cat_interest_2}%
  \BibitemOpen
  \bibfield  {author} {\bibinfo {author} {\bibfnamefont {W.}~\bibnamefont
  {Qin}}, \bibinfo {author} {\bibfnamefont {A.}~\bibnamefont {Miranowicz}},
  \bibinfo {author} {\bibfnamefont {H.}~\bibnamefont {Jing}},\ and\ \bibinfo
  {author} {\bibfnamefont {F.}~\bibnamefont {Nori}},\ }\bibfield  {title}
  {\bibinfo {title} {Generating long-lived macroscopically distinct
  superposition states in atomic ensembles},\ }\href
  {https://doi.org/10.1103/PhysRevLett.127.093602} {\bibfield  {journal}
  {\bibinfo  {journal} {Phys. Rev. Lett.}\ }\textbf {\bibinfo {volume} {127}},\
  \bibinfo {pages} {093602} (\bibinfo {year} {2021})}\BibitemShut {NoStop}%
\bibitem [{\citenamefont {Kang}\ \emph {et~al.}(2022)\citenamefont {Kang},
  \citenamefont {Chen}, \citenamefont {Wang}, \citenamefont {Song},
  \citenamefont {Xia}, \citenamefont {Miranowicz}, \citenamefont {Zheng},\ and\
  \citenamefont {Nori}}]{cat_interest_3}%
  \BibitemOpen
  \bibfield  {author} {\bibinfo {author} {\bibfnamefont {Y.-H.}\ \bibnamefont
  {Kang}}, \bibinfo {author} {\bibfnamefont {Y.-H.}\ \bibnamefont {Chen}},
  \bibinfo {author} {\bibfnamefont {X.}~\bibnamefont {Wang}}, \bibinfo {author}
  {\bibfnamefont {J.}~\bibnamefont {Song}}, \bibinfo {author} {\bibfnamefont
  {Y.}~\bibnamefont {Xia}}, \bibinfo {author} {\bibfnamefont {A.}~\bibnamefont
  {Miranowicz}}, \bibinfo {author} {\bibfnamefont {S.-B.}\ \bibnamefont
  {Zheng}},\ and\ \bibinfo {author} {\bibfnamefont {F.}~\bibnamefont {Nori}},\
  }\bibfield  {title} {\bibinfo {title} {Nonadiabatic geometric quantum
  computation with cat-state qubits via invariant-based reverse engineering},\
  }\href {https://doi.org/10.1103/PhysRevResearch.4.013233} {\bibfield
  {journal} {\bibinfo  {journal} {Phys. Rev. Res.}\ }\textbf {\bibinfo {volume}
  {4}},\ \bibinfo {pages} {013233} (\bibinfo {year} {2022})}\BibitemShut
  {NoStop}%
\bibitem [{\citenamefont {Omran}\ \emph {et~al.}(2019)\citenamefont {Omran},
  \citenamefont {Levine}, \citenamefont {Keesling}, \citenamefont {Semeghini},
  \citenamefont {Wang}, \citenamefont {Ebadi}, \citenamefont {Bernien},
  \citenamefont {Zibrov}, \citenamefont {Pichler}, \citenamefont {Choi},
  \citenamefont {Cui}, \citenamefont {Rossignolo}, \citenamefont {Rembold},
  \citenamefont {Montangero}, \citenamefont {Calarco}, \citenamefont {Endres},
  \citenamefont {Greiner}, \citenamefont {Vuleti{\'{c}}},\ and\ \citenamefont
  {Lukin}}]{lukinScience}%
  \BibitemOpen
  \bibfield  {author} {\bibinfo {author} {\bibfnamefont {A.}~\bibnamefont
  {Omran}}, \bibinfo {author} {\bibfnamefont {H.}~\bibnamefont {Levine}},
  \bibinfo {author} {\bibfnamefont {A.}~\bibnamefont {Keesling}}, \bibinfo
  {author} {\bibfnamefont {G.}~\bibnamefont {Semeghini}}, \bibinfo {author}
  {\bibfnamefont {T.~T.}\ \bibnamefont {Wang}}, \bibinfo {author}
  {\bibfnamefont {S.}~\bibnamefont {Ebadi}}, \bibinfo {author} {\bibfnamefont
  {H.}~\bibnamefont {Bernien}}, \bibinfo {author} {\bibfnamefont {A.~S.}\
  \bibnamefont {Zibrov}}, \bibinfo {author} {\bibfnamefont {H.}~\bibnamefont
  {Pichler}}, \bibinfo {author} {\bibfnamefont {S.}~\bibnamefont {Choi}},
  \bibinfo {author} {\bibfnamefont {J.}~\bibnamefont {Cui}}, \bibinfo {author}
  {\bibfnamefont {M.}~\bibnamefont {Rossignolo}}, \bibinfo {author}
  {\bibfnamefont {P.}~\bibnamefont {Rembold}}, \bibinfo {author} {\bibfnamefont
  {S.}~\bibnamefont {Montangero}}, \bibinfo {author} {\bibfnamefont
  {T.}~\bibnamefont {Calarco}}, \bibinfo {author} {\bibfnamefont
  {M.}~\bibnamefont {Endres}}, \bibinfo {author} {\bibfnamefont
  {M.}~\bibnamefont {Greiner}}, \bibinfo {author} {\bibfnamefont
  {V.}~\bibnamefont {Vuleti{\'{c}}}},\ and\ \bibinfo {author} {\bibfnamefont
  {M.~D.}\ \bibnamefont {Lukin}},\ }\bibfield  {title} {\bibinfo {title}
  {Generation and manipulation of {Schr\"{o}dinger} cat states in {Rydberg}
  atom arrays},\ }\href {https://doi.org/10.1126/science.aax9743} {\bibfield
  {journal} {\bibinfo  {journal} {Science}\ }\textbf {\bibinfo {volume}
  {365}},\ \bibinfo {pages} {570} (\bibinfo {year} {2019})}\BibitemShut
  {NoStop}%
\bibitem [{\citenamefont {Song}\ \emph {et~al.}(2019)\citenamefont {Song},
  \citenamefont {Xu}, \citenamefont {Li}, \citenamefont {Zhang}, \citenamefont
  {Zhang}, \citenamefont {Liu}, \citenamefont {Guo}, \citenamefont {Wang},
  \citenamefont {Ren}, \citenamefont {Hao}, \citenamefont {Feng}, \citenamefont
  {Fan}, \citenamefont {Zheng}, \citenamefont {Wang}, \citenamefont {Wang},\
  and\ \citenamefont {Zhu}}]{songcat}%
  \BibitemOpen
  \bibfield  {author} {\bibinfo {author} {\bibfnamefont {C.}~\bibnamefont
  {Song}}, \bibinfo {author} {\bibfnamefont {K.}~\bibnamefont {Xu}}, \bibinfo
  {author} {\bibfnamefont {H.}~\bibnamefont {Li}}, \bibinfo {author}
  {\bibfnamefont {Y.-R.}\ \bibnamefont {Zhang}}, \bibinfo {author}
  {\bibfnamefont {X.}~\bibnamefont {Zhang}}, \bibinfo {author} {\bibfnamefont
  {W.}~\bibnamefont {Liu}}, \bibinfo {author} {\bibfnamefont {Q.}~\bibnamefont
  {Guo}}, \bibinfo {author} {\bibfnamefont {Z.}~\bibnamefont {Wang}}, \bibinfo
  {author} {\bibfnamefont {W.}~\bibnamefont {Ren}}, \bibinfo {author}
  {\bibfnamefont {J.}~\bibnamefont {Hao}}, \bibinfo {author} {\bibfnamefont
  {H.}~\bibnamefont {Feng}}, \bibinfo {author} {\bibfnamefont {H.}~\bibnamefont
  {Fan}}, \bibinfo {author} {\bibfnamefont {D.}~\bibnamefont {Zheng}}, \bibinfo
  {author} {\bibfnamefont {D.-W.}\ \bibnamefont {Wang}}, \bibinfo {author}
  {\bibfnamefont {H.}~\bibnamefont {Wang}},\ and\ \bibinfo {author}
  {\bibfnamefont {S.-Y.}\ \bibnamefont {Zhu}},\ }\bibfield  {title} {\bibinfo
  {title} {{Generation of multicomponent atomic {Schr\"{o}dinger} cat states of
  up to 20 qubits}},\ }\href {https://doi.org/10.1126/science.aay0600}
  {\bibfield  {journal} {\bibinfo  {journal} {Science}\ }\textbf {\bibinfo
  {volume} {365}},\ \bibinfo {pages} {574} (\bibinfo {year}
  {2019})}\BibitemShut {NoStop}%
\bibitem [{\citenamefont {Ravets}\ \emph {et~al.}(2014)\citenamefont {Ravets},
  \citenamefont {Labuhn}, \citenamefont {Barredo}, \citenamefont
  {B{\'{e}}guin}, \citenamefont {Lahaye},\ and\ \citenamefont
  {Browaeys}}]{Forster}%
  \BibitemOpen
  \bibfield  {author} {\bibinfo {author} {\bibfnamefont {S.}~\bibnamefont
  {Ravets}}, \bibinfo {author} {\bibfnamefont {H.}~\bibnamefont {Labuhn}},
  \bibinfo {author} {\bibfnamefont {D.}~\bibnamefont {Barredo}}, \bibinfo
  {author} {\bibfnamefont {L.}~\bibnamefont {B{\'{e}}guin}}, \bibinfo {author}
  {\bibfnamefont {T.}~\bibnamefont {Lahaye}},\ and\ \bibinfo {author}
  {\bibfnamefont {A.}~\bibnamefont {Browaeys}},\ }\bibfield  {title} {\bibinfo
  {title} {Coherent dipole{\textendash}dipole coupling between two single
  {Rydberg} atoms at an electrically-tuned {F\"{o}rster} resonance},\ }\href
  {https://doi.org/10.1038/nphys3119} {\bibfield  {journal} {\bibinfo
  {journal} {Nat. Phys.}\ }\textbf {\bibinfo {volume} {10}},\ \bibinfo {pages}
  {914} (\bibinfo {year} {2014})}\BibitemShut {NoStop}%
\bibitem [{\citenamefont {Madjarov}\ \emph {et~al.}(2020)\citenamefont
  {Madjarov}, \citenamefont {Covey}, \citenamefont {Shaw}, \citenamefont
  {Choi}, \citenamefont {Kale}, \citenamefont {Cooper}, \citenamefont
  {Pichler}, \citenamefont {Schkolnik}, \citenamefont {Williams},\ and\
  \citenamefont {Endres}}]{highfideliyNP}%
  \BibitemOpen
  \bibfield  {author} {\bibinfo {author} {\bibfnamefont {I.~S.}\ \bibnamefont
  {Madjarov}}, \bibinfo {author} {\bibfnamefont {J.~P.}\ \bibnamefont {Covey}},
  \bibinfo {author} {\bibfnamefont {A.~L.}\ \bibnamefont {Shaw}}, \bibinfo
  {author} {\bibfnamefont {J.}~\bibnamefont {Choi}}, \bibinfo {author}
  {\bibfnamefont {A.}~\bibnamefont {Kale}}, \bibinfo {author} {\bibfnamefont
  {A.}~\bibnamefont {Cooper}}, \bibinfo {author} {\bibfnamefont
  {H.}~\bibnamefont {Pichler}}, \bibinfo {author} {\bibfnamefont
  {V.}~\bibnamefont {Schkolnik}}, \bibinfo {author} {\bibfnamefont {J.~R.}\
  \bibnamefont {Williams}},\ and\ \bibinfo {author} {\bibfnamefont
  {M.}~\bibnamefont {Endres}},\ }\bibfield  {title} {\bibinfo {title}
  {{High-fidelity entanglement and detection of alkaline-earth Rydberg
  atoms}},\ }\href {https://doi.org/10.1038/s41567-020-0903-z} {\bibfield
  {journal} {\bibinfo  {journal} {Nat. Phys.}\ }\textbf {\bibinfo {volume}
  {16}},\ \bibinfo {pages} {857} (\bibinfo {year} {2020})}\BibitemShut
  {NoStop}%
\bibitem [{\citenamefont {Kang}\ \emph {et~al.}(2018)\citenamefont {Kang},
  \citenamefont {Chen}, \citenamefont {Shi}, \citenamefont {Huang},
  \citenamefont {Song},\ and\ \citenamefont {Xia}}]{kang1}%
  \BibitemOpen
  \bibfield  {author} {\bibinfo {author} {\bibfnamefont {Y.-H.}\ \bibnamefont
  {Kang}}, \bibinfo {author} {\bibfnamefont {Y.-H.}\ \bibnamefont {Chen}},
  \bibinfo {author} {\bibfnamefont {Z.-C.}\ \bibnamefont {Shi}}, \bibinfo
  {author} {\bibfnamefont {B.-H.}\ \bibnamefont {Huang}}, \bibinfo {author}
  {\bibfnamefont {J.}~\bibnamefont {Song}},\ and\ \bibinfo {author}
  {\bibfnamefont {Y.}~\bibnamefont {Xia}},\ }\bibfield  {title} {\bibinfo
  {title} {{Nonadiabatic holonomic quantum computation using Rydberg
  blockade}},\ }\href {https://doi.org/10.1103/PhysRevA.97.042336} {\bibfield
  {journal} {\bibinfo  {journal} {Phys. Rev. A}\ }\textbf {\bibinfo {volume}
  {97}},\ \bibinfo {pages} {042336} (\bibinfo {year} {2018})}\BibitemShut
  {NoStop}%
\bibitem [{\citenamefont {Wang}\ \emph {et~al.}(2019)\citenamefont {Wang},
  \citenamefont {Hu}, \citenamefont {Shi}, \citenamefont {Huang}, \citenamefont
  {Song},\ and\ \citenamefont {Xia}}]{wangyu}%
  \BibitemOpen
  \bibfield  {author} {\bibinfo {author} {\bibfnamefont {Y.}~\bibnamefont
  {Wang}}, \bibinfo {author} {\bibfnamefont {C.-S.}\ \bibnamefont {Hu}},
  \bibinfo {author} {\bibfnamefont {Z.-C.}\ \bibnamefont {Shi}}, \bibinfo
  {author} {\bibfnamefont {B.-H.}\ \bibnamefont {Huang}}, \bibinfo {author}
  {\bibfnamefont {J.}~\bibnamefont {Song}},\ and\ \bibinfo {author}
  {\bibfnamefont {Y.}~\bibnamefont {Xia}},\ }\bibfield  {title} {\bibinfo
  {title} {Accelerated and noise-resistant protocol of dissipation-based
  {Knill–Laflamme–Milburn} state generation with {Lyapunov} control},\
  }\href {https://doi.org/https://doi.org/10.1002/andp.201900006} {\bibfield
  {journal} {\bibinfo  {journal} {Ann. Phys. (Berlin)}\ }\textbf {\bibinfo
  {volume} {531}},\ \bibinfo {pages} {1900006} (\bibinfo {year}
  {2019})}\BibitemShut {NoStop}%
\bibitem [{\citenamefont {Su}\ \emph {et~al.}(2020)\citenamefont {Su},
  \citenamefont {Guo}, \citenamefont {Tian}, \citenamefont {Zhu}, \citenamefont
  {Yan}, \citenamefont {Liang},\ and\ \citenamefont {Feng}}]{su1}%
  \BibitemOpen
  \bibfield  {author} {\bibinfo {author} {\bibfnamefont {S.-L.}\ \bibnamefont
  {Su}}, \bibinfo {author} {\bibfnamefont {F.-Q.}\ \bibnamefont {Guo}},
  \bibinfo {author} {\bibfnamefont {L.}~\bibnamefont {Tian}}, \bibinfo {author}
  {\bibfnamefont {X.-Y.}\ \bibnamefont {Zhu}}, \bibinfo {author} {\bibfnamefont
  {L.-L.}\ \bibnamefont {Yan}}, \bibinfo {author} {\bibfnamefont {E.-J.}\
  \bibnamefont {Liang}},\ and\ \bibinfo {author} {\bibfnamefont
  {M.}~\bibnamefont {Feng}},\ }\bibfield  {title} {\bibinfo {title}
  {{Nondestructive Rydberg parity meter and its applications}},\ }\href
  {https://doi.org/10.1103/PhysRevA.101.012347} {\bibfield  {journal} {\bibinfo
   {journal} {Phys. Rev. A}\ }\textbf {\bibinfo {volume} {101}},\ \bibinfo
  {pages} {012347} (\bibinfo {year} {2020})}\BibitemShut {NoStop}%
\bibitem [{\citenamefont {Zheng}\ \emph {et~al.}(2020)\citenamefont {Zheng},
  \citenamefont {Kang}, \citenamefont {Ran}, \citenamefont {Shi},\ and\
  \citenamefont {Xia}}]{zheng1}%
  \BibitemOpen
  \bibfield  {author} {\bibinfo {author} {\bibfnamefont {R.-H.}\ \bibnamefont
  {Zheng}}, \bibinfo {author} {\bibfnamefont {Y.-H.}\ \bibnamefont {Kang}},
  \bibinfo {author} {\bibfnamefont {D.}~\bibnamefont {Ran}}, \bibinfo {author}
  {\bibfnamefont {Z.-C.}\ \bibnamefont {Shi}},\ and\ \bibinfo {author}
  {\bibfnamefont {Y.}~\bibnamefont {Xia}},\ }\bibfield  {title} {\bibinfo
  {title} {{Deterministic interconversions between the
  Greenberger-Horne-Zeilinger states and the $W$ states by invariant-based
  pulse design}},\ }\href {https://doi.org/10.1103/PhysRevA.101.012345}
  {\bibfield  {journal} {\bibinfo  {journal} {Phys. Rev. A}\ }\textbf {\bibinfo
  {volume} {101}},\ \bibinfo {pages} {012345} (\bibinfo {year}
  {2020})}\BibitemShut {NoStop}%
\bibitem [{\citenamefont {Kang}\ \emph {et~al.}(2020)\citenamefont {Kang},
  \citenamefont {Shi}, \citenamefont {Song},\ and\ \citenamefont
  {Xia}}]{kang2}%
  \BibitemOpen
  \bibfield  {author} {\bibinfo {author} {\bibfnamefont {Y.-H.}\ \bibnamefont
  {Kang}}, \bibinfo {author} {\bibfnamefont {Z.-C.}\ \bibnamefont {Shi}},
  \bibinfo {author} {\bibfnamefont {J.}~\bibnamefont {Song}},\ and\ \bibinfo
  {author} {\bibfnamefont {Y.}~\bibnamefont {Xia}},\ }\bibfield  {title}
  {\bibinfo {title} {{Heralded atomic nonadiabatic holonomic quantum
  computation with Rydberg blockade}},\ }\href
  {https://doi.org/10.1103/PhysRevA.102.022617} {\bibfield  {journal} {\bibinfo
   {journal} {Phys. Rev. A}\ }\textbf {\bibinfo {volume} {102}},\ \bibinfo
  {pages} {022617} (\bibinfo {year} {2020})}\BibitemShut {NoStop}%
\bibitem [{\citenamefont {Su}\ and\ \citenamefont {Li}(2021)}]{su2}%
  \BibitemOpen
  \bibfield  {author} {\bibinfo {author} {\bibfnamefont {S.-L.}\ \bibnamefont
  {Su}}\ and\ \bibinfo {author} {\bibfnamefont {W.}~\bibnamefont {Li}},\
  }\bibfield  {title} {\bibinfo {title} {{Dipole-dipole-interaction--driven
  antiblockade of two Rydberg atoms}},\ }\href
  {https://doi.org/10.1103/PhysRevA.104.033716} {\bibfield  {journal} {\bibinfo
   {journal} {Phys. Rev. A}\ }\textbf {\bibinfo {volume} {104}},\ \bibinfo
  {pages} {033716} (\bibinfo {year} {2021})}\BibitemShut {NoStop}%
\bibitem [{\citenamefont {Liu}\ \emph {et~al.}(2022)\citenamefont {Liu},
  \citenamefont {Shen}, \citenamefont {Zheng}, \citenamefont {Kang},
  \citenamefont {Shi}, \citenamefont {Song},\ and\ \citenamefont
  {Xia}}]{liushuai}%
  \BibitemOpen
  \bibfield  {author} {\bibinfo {author} {\bibfnamefont {S.}~\bibnamefont
  {Liu}}, \bibinfo {author} {\bibfnamefont {J.-H.}\ \bibnamefont {Shen}},
  \bibinfo {author} {\bibfnamefont {R.-H.}\ \bibnamefont {Zheng}}, \bibinfo
  {author} {\bibfnamefont {Y.-H.}\ \bibnamefont {Kang}}, \bibinfo {author}
  {\bibfnamefont {Z.-C.}\ \bibnamefont {Shi}}, \bibinfo {author} {\bibfnamefont
  {J.}~\bibnamefont {Song}},\ and\ \bibinfo {author} {\bibfnamefont
  {Y.}~\bibnamefont {Xia}},\ }\bibfield  {title} {\bibinfo {title} {Optimized
  nonadiabatic holonomic quantum computation based on {F\"orster} resonance in
  {Rydberg} atoms},\ }\href {https://doi.org/10.1007/s11467-021-1108-3}
  {\bibfield  {journal} {\bibinfo  {journal} {Frontiers of Physics}\ }\textbf
  {\bibinfo {volume} {17}},\ \bibinfo {pages} {21502} (\bibinfo {year}
  {2022})}\BibitemShut {NoStop}%
\bibitem [{\citenamefont {Levine}\ \emph {et~al.}(2018)\citenamefont {Levine},
  \citenamefont {Keesling}, \citenamefont {Omran}, \citenamefont {Bernien},
  \citenamefont {Schwartz}, \citenamefont {Zibrov}, \citenamefont {Endres},
  \citenamefont {Greiner}, \citenamefont {Vuleti\ifmmode~\acute{c}\else
  \'{c}\fi{}},\ and\ \citenamefont {Lukin}}]{Dopplerdata-PRL}%
  \BibitemOpen
  \bibfield  {author} {\bibinfo {author} {\bibfnamefont {H.}~\bibnamefont
  {Levine}}, \bibinfo {author} {\bibfnamefont {A.}~\bibnamefont {Keesling}},
  \bibinfo {author} {\bibfnamefont {A.}~\bibnamefont {Omran}}, \bibinfo
  {author} {\bibfnamefont {H.}~\bibnamefont {Bernien}}, \bibinfo {author}
  {\bibfnamefont {S.}~\bibnamefont {Schwartz}}, \bibinfo {author}
  {\bibfnamefont {A.~S.}\ \bibnamefont {Zibrov}}, \bibinfo {author}
  {\bibfnamefont {M.}~\bibnamefont {Endres}}, \bibinfo {author} {\bibfnamefont
  {M.}~\bibnamefont {Greiner}}, \bibinfo {author} {\bibfnamefont
  {V.}~\bibnamefont {Vuleti\ifmmode~\acute{c}\else \'{c}\fi{}}},\ and\ \bibinfo
  {author} {\bibfnamefont {M.~D.}\ \bibnamefont {Lukin}},\ }\bibfield  {title}
  {\bibinfo {title} {High-fidelity control and entanglement of {Rydberg}-atom
  qubits},\ }\href {https://doi.org/10.1103/PhysRevLett.121.123603} {\bibfield
  {journal} {\bibinfo  {journal} {Phys. Rev. Lett.}\ }\textbf {\bibinfo
  {volume} {121}},\ \bibinfo {pages} {123603} (\bibinfo {year}
  {2018})}\BibitemShut {NoStop}%
\bibitem [{\citenamefont {Degen}\ \emph {et~al.}(2017)\citenamefont {Degen},
  \citenamefont {Reinhard},\ and\ \citenamefont {Cappellaro}}]{QS-GHZ}%
  \BibitemOpen
  \bibfield  {author} {\bibinfo {author} {\bibfnamefont {C.~L.}\ \bibnamefont
  {Degen}}, \bibinfo {author} {\bibfnamefont {F.}~\bibnamefont {Reinhard}},\
  and\ \bibinfo {author} {\bibfnamefont {P.}~\bibnamefont {Cappellaro}},\
  }\bibfield  {title} {\bibinfo {title} {Quantum sensing},\ }\href
  {https://doi.org/10.1103/RevModPhys.89.035002} {\bibfield  {journal}
  {\bibinfo  {journal} {Rev. Mod. Phys.}\ }\textbf {\bibinfo {volume} {89}},\
  \bibinfo {pages} {035002} (\bibinfo {year} {2017})}\BibitemShut {NoStop}%
\bibitem [{\citenamefont {D\"ur}\ \emph {et~al.}(2014)\citenamefont {D\"ur},
  \citenamefont {Skotiniotis}, \citenamefont {Fr\"owis},\ and\ \citenamefont
  {Kraus}}]{QM-GHZ}%
  \BibitemOpen
  \bibfield  {author} {\bibinfo {author} {\bibfnamefont {W.}~\bibnamefont
  {D\"ur}}, \bibinfo {author} {\bibfnamefont {M.}~\bibnamefont {Skotiniotis}},
  \bibinfo {author} {\bibfnamefont {F.}~\bibnamefont {Fr\"owis}},\ and\
  \bibinfo {author} {\bibfnamefont {B.}~\bibnamefont {Kraus}},\ }\bibfield
  {title} {\bibinfo {title} {Improved quantum metrology using quantum error
  correction},\ }\href {https://doi.org/10.1103/PhysRevLett.112.080801}
  {\bibfield  {journal} {\bibinfo  {journal} {Phys. Rev. Lett.}\ }\textbf
  {\bibinfo {volume} {112}},\ \bibinfo {pages} {080801} (\bibinfo {year}
  {2014})}\BibitemShut {NoStop}%
\bibitem [{\citenamefont {Egan}\ \emph {et~al.}()\citenamefont {Egan},
  \citenamefont {Debroy}, \citenamefont {Noel}, \citenamefont {Risinger},
  \citenamefont {Zhu}, \citenamefont {Biswas}, \citenamefont {Newman},
  \citenamefont {Li}, \citenamefont {Brown}, \citenamefont {Cetina},\ and\
  \citenamefont {Monroe}}]{QEC-GHZ}%
  \BibitemOpen
  \bibfield  {author} {\bibinfo {author} {\bibfnamefont {L.}~\bibnamefont
  {Egan}}, \bibinfo {author} {\bibfnamefont {D.~M.}\ \bibnamefont {Debroy}},
  \bibinfo {author} {\bibfnamefont {C.}~\bibnamefont {Noel}}, \bibinfo {author}
  {\bibfnamefont {A.}~\bibnamefont {Risinger}}, \bibinfo {author}
  {\bibfnamefont {D.}~\bibnamefont {Zhu}}, \bibinfo {author} {\bibfnamefont
  {D.}~\bibnamefont {Biswas}}, \bibinfo {author} {\bibfnamefont
  {M.}~\bibnamefont {Newman}}, \bibinfo {author} {\bibfnamefont
  {M.}~\bibnamefont {Li}}, \bibinfo {author} {\bibfnamefont {K.~R.}\
  \bibnamefont {Brown}}, \bibinfo {author} {\bibfnamefont {M.}~\bibnamefont
  {Cetina}},\ and\ \bibinfo {author} {\bibfnamefont {C.}~\bibnamefont
  {Monroe}},\ }\href@noop {} {\bibinfo {title} {Fault-tolerant operation of a
  quantum error-correction code}},\ \bibinfo {note} {arXiv:2009.11482v2
  (2020)}\BibitemShut {NoStop}%
\bibitem [{\citenamefont {Wilk}\ \emph {et~al.}(2010)\citenamefont {Wilk},
  \citenamefont {Ga\"etan}, \citenamefont {Evellin}, \citenamefont {Wolters},
  \citenamefont {Miroshnychenko}, \citenamefont {Grangier},\ and\ \citenamefont
  {Browaeys}}]{lukinPRL}%
  \BibitemOpen
  \bibfield  {author} {\bibinfo {author} {\bibfnamefont {T.}~\bibnamefont
  {Wilk}}, \bibinfo {author} {\bibfnamefont {A.}~\bibnamefont {Ga\"etan}},
  \bibinfo {author} {\bibfnamefont {C.}~\bibnamefont {Evellin}}, \bibinfo
  {author} {\bibfnamefont {J.}~\bibnamefont {Wolters}}, \bibinfo {author}
  {\bibfnamefont {Y.}~\bibnamefont {Miroshnychenko}}, \bibinfo {author}
  {\bibfnamefont {P.}~\bibnamefont {Grangier}},\ and\ \bibinfo {author}
  {\bibfnamefont {A.}~\bibnamefont {Browaeys}},\ }\bibfield  {title} {\bibinfo
  {title} {Entanglement of two individual neutral atoms using {Rydberg}
  blockade},\ }\href {https://doi.org/10.1103/PhysRevLett.104.010502}
  {\bibfield  {journal} {\bibinfo  {journal} {Phys. Rev. Lett.}\ }\textbf
  {\bibinfo {volume} {104}},\ \bibinfo {pages} {010502} (\bibinfo {year}
  {2010})}\BibitemShut {NoStop}%
\bibitem [{\citenamefont {Ates}\ \emph {et~al.}(2007)\citenamefont {Ates},
  \citenamefont {Pohl}, \citenamefont {Pattard},\ and\ \citenamefont
  {Rost}}]{CTPRL98}%
  \BibitemOpen
  \bibfield  {author} {\bibinfo {author} {\bibfnamefont {C.}~\bibnamefont
  {Ates}}, \bibinfo {author} {\bibfnamefont {T.}~\bibnamefont {Pohl}}, \bibinfo
  {author} {\bibfnamefont {T.}~\bibnamefont {Pattard}},\ and\ \bibinfo {author}
  {\bibfnamefont {J.~M.}\ \bibnamefont {Rost}},\ }\bibfield  {title} {\bibinfo
  {title} {Antiblockade in {Rydberg} excitation of an ultracold lattice gas},\
  }\href {https://doi.org/10.1103/PhysRevLett.98.023002} {\bibfield  {journal}
  {\bibinfo  {journal} {Phys. Rev. Lett.}\ }\textbf {\bibinfo {volume} {98}},\
  \bibinfo {pages} {023002} (\bibinfo {year} {2007})}\BibitemShut {NoStop}%
\bibitem [{\citenamefont {Amthor}\ \emph {et~al.}(2010)\citenamefont {Amthor},
  \citenamefont {Giese}, \citenamefont {Hofmann},\ and\ \citenamefont
  {Weidem\"uller}}]{TCPRL104}%
  \BibitemOpen
  \bibfield  {author} {\bibinfo {author} {\bibfnamefont {T.}~\bibnamefont
  {Amthor}}, \bibinfo {author} {\bibfnamefont {C.}~\bibnamefont {Giese}},
  \bibinfo {author} {\bibfnamefont {C.~S.}\ \bibnamefont {Hofmann}},\ and\
  \bibinfo {author} {\bibfnamefont {M.}~\bibnamefont {Weidem\"uller}},\
  }\bibfield  {title} {\bibinfo {title} {Evidence of antiblockade in an
  ultracold {Rydberg} gas},\ }\href
  {https://doi.org/10.1103/PhysRevLett.104.013001} {\bibfield  {journal}
  {\bibinfo  {journal} {Phys. Rev. Lett.}\ }\textbf {\bibinfo {volume} {104}},\
  \bibinfo {pages} {013001} (\bibinfo {year} {2010})}\BibitemShut {NoStop}%
\bibitem [{\citenamefont {Zuo}\ and\ \citenamefont {Nakagawa}(2010)}]{ZKPRA82}%
  \BibitemOpen
  \bibfield  {author} {\bibinfo {author} {\bibfnamefont {Z.}~\bibnamefont
  {Zuo}}\ and\ \bibinfo {author} {\bibfnamefont {K.}~\bibnamefont {Nakagawa}},\
  }\bibfield  {title} {\bibinfo {title} {Multiparticle entanglement in a
  one-dimensional optical lattice using {Rydberg}-atom interactions},\ }\href
  {https://doi.org/10.1103/PhysRevA.82.062328} {\bibfield  {journal} {\bibinfo
  {journal} {Phys. Rev. A}\ }\textbf {\bibinfo {volume} {82}},\ \bibinfo
  {pages} {062328} (\bibinfo {year} {2010})}\BibitemShut {NoStop}%
\bibitem [{\citenamefont {Lee}\ \emph {et~al.}(2012)\citenamefont {Lee},
  \citenamefont {H\"affner},\ and\ \citenamefont {Cross}}]{THPRL108}%
  \BibitemOpen
  \bibfield  {author} {\bibinfo {author} {\bibfnamefont {T.~E.}\ \bibnamefont
  {Lee}}, \bibinfo {author} {\bibfnamefont {H.}~\bibnamefont {H\"affner}},\
  and\ \bibinfo {author} {\bibfnamefont {M.~C.}\ \bibnamefont {Cross}},\
  }\bibfield  {title} {\bibinfo {title} {Collective quantum jumps of {Rydberg}
  atoms},\ }\href {https://doi.org/10.1103/PhysRevLett.108.023602} {\bibfield
  {journal} {\bibinfo  {journal} {Phys. Rev. Lett.}\ }\textbf {\bibinfo
  {volume} {108}},\ \bibinfo {pages} {023602} (\bibinfo {year}
  {2012})}\BibitemShut {NoStop}%
\bibitem [{\citenamefont {Li}\ \emph {et~al.}(2013)\citenamefont {Li},
  \citenamefont {Ates},\ and\ \citenamefont {Lesanovsky}}]{WCPRL110}%
  \BibitemOpen
  \bibfield  {author} {\bibinfo {author} {\bibfnamefont {W.}~\bibnamefont
  {Li}}, \bibinfo {author} {\bibfnamefont {C.}~\bibnamefont {Ates}},\ and\
  \bibinfo {author} {\bibfnamefont {I.}~\bibnamefont {Lesanovsky}},\ }\bibfield
   {title} {\bibinfo {title} {Nonadiabatic motional effects and dissipative
  blockade for {Rydberg} atoms excited from optical lattices or microtraps},\
  }\href {https://doi.org/10.1103/PhysRevLett.110.213005} {\bibfield  {journal}
  {\bibinfo  {journal} {Phys. Rev. Lett.}\ }\textbf {\bibinfo {volume} {110}},\
  \bibinfo {pages} {213005} (\bibinfo {year} {2013})}\BibitemShut {NoStop}%
\bibitem [{\citenamefont {Su}\ \emph {et~al.}(2016)\citenamefont {Su},
  \citenamefont {Liang}, \citenamefont {Zhang}, \citenamefont {Wen},
  \citenamefont {Sun}, \citenamefont {Jin},\ and\ \citenamefont
  {Zhu}}]{SEPRA93}%
  \BibitemOpen
  \bibfield  {author} {\bibinfo {author} {\bibfnamefont {S.-L.}\ \bibnamefont
  {Su}}, \bibinfo {author} {\bibfnamefont {E.}~\bibnamefont {Liang}}, \bibinfo
  {author} {\bibfnamefont {S.}~\bibnamefont {Zhang}}, \bibinfo {author}
  {\bibfnamefont {J.-J.}\ \bibnamefont {Wen}}, \bibinfo {author} {\bibfnamefont
  {L.-L.}\ \bibnamefont {Sun}}, \bibinfo {author} {\bibfnamefont
  {Z.}~\bibnamefont {Jin}},\ and\ \bibinfo {author} {\bibfnamefont {A.-D.}\
  \bibnamefont {Zhu}},\ }\bibfield  {title} {\bibinfo {title} {One-step
  implementation of the {Rydberg-Rydberg-interaction} gate},\ }\href
  {https://doi.org/10.1103/PhysRevA.93.012306} {\bibfield  {journal} {\bibinfo
  {journal} {Phys. Rev. A}\ }\textbf {\bibinfo {volume} {93}},\ \bibinfo
  {pages} {012306} (\bibinfo {year} {2016})}\BibitemShut {NoStop}%
\bibitem [{\citenamefont {Rach}\ \emph {et~al.}(2015)\citenamefont {Rach},
  \citenamefont {M\"uller}, \citenamefont {Calarco},\ and\ \citenamefont
  {Montangero}}]{OPT1}%
  \BibitemOpen
  \bibfield  {author} {\bibinfo {author} {\bibfnamefont {N.}~\bibnamefont
  {Rach}}, \bibinfo {author} {\bibfnamefont {M.~M.}\ \bibnamefont {M\"uller}},
  \bibinfo {author} {\bibfnamefont {T.}~\bibnamefont {Calarco}},\ and\ \bibinfo
  {author} {\bibfnamefont {S.}~\bibnamefont {Montangero}},\ }\bibfield  {title}
  {\bibinfo {title} {{Dressing the chopped-random-basis optimization: {A}
  bandwidth-limited access to the trap-free landscape}},\ }\href
  {https://doi.org/10.1103/PhysRevA.92.062343} {\bibfield  {journal} {\bibinfo
  {journal} {Phys. Rev. A}\ }\textbf {\bibinfo {volume} {92}},\ \bibinfo
  {pages} {062343} (\bibinfo {year} {2015})}\BibitemShut {NoStop}%
\bibitem [{\citenamefont {Heck}\ \emph {et~al.}(2018)\citenamefont {Heck},
  \citenamefont {Vuculescu}, \citenamefont {S{\o}rensen}, \citenamefont
  {Zoller}, \citenamefont {Andreasen}, \citenamefont {Bason}, \citenamefont
  {Ejlertsen}, \citenamefont {El{\'{\i}}asson}, \citenamefont {Haikka},
  \citenamefont {Laustsen}, \citenamefont {Nielsen}, \citenamefont {Mao},
  \citenamefont {M\"{u}ller}, \citenamefont {Napolitano}, \citenamefont
  {Pedersen}, \citenamefont {Thorsen}, \citenamefont {Bergenholtz},
  \citenamefont {Calarco}, \citenamefont {Montangero},\ and\ \citenamefont
  {Sherson}}]{OPT2}%
  \BibitemOpen
  \bibfield  {author} {\bibinfo {author} {\bibfnamefont {R.}~\bibnamefont
  {Heck}}, \bibinfo {author} {\bibfnamefont {O.}~\bibnamefont {Vuculescu}},
  \bibinfo {author} {\bibfnamefont {J.~J.}\ \bibnamefont {S{\o}rensen}},
  \bibinfo {author} {\bibfnamefont {J.}~\bibnamefont {Zoller}}, \bibinfo
  {author} {\bibfnamefont {M.~G.}\ \bibnamefont {Andreasen}}, \bibinfo {author}
  {\bibfnamefont {M.~G.}\ \bibnamefont {Bason}}, \bibinfo {author}
  {\bibfnamefont {P.}~\bibnamefont {Ejlertsen}}, \bibinfo {author}
  {\bibfnamefont {O.}~\bibnamefont {El{\'{\i}}asson}}, \bibinfo {author}
  {\bibfnamefont {P.}~\bibnamefont {Haikka}}, \bibinfo {author} {\bibfnamefont
  {J.~S.}\ \bibnamefont {Laustsen}}, \bibinfo {author} {\bibfnamefont {L.~L.}\
  \bibnamefont {Nielsen}}, \bibinfo {author} {\bibfnamefont {A.}~\bibnamefont
  {Mao}}, \bibinfo {author} {\bibfnamefont {R.}~\bibnamefont {M\"{u}ller}},
  \bibinfo {author} {\bibfnamefont {M.}~\bibnamefont {Napolitano}}, \bibinfo
  {author} {\bibfnamefont {M.~K.}\ \bibnamefont {Pedersen}}, \bibinfo {author}
  {\bibfnamefont {A.~R.}\ \bibnamefont {Thorsen}}, \bibinfo {author}
  {\bibfnamefont {C.}~\bibnamefont {Bergenholtz}}, \bibinfo {author}
  {\bibfnamefont {T.}~\bibnamefont {Calarco}}, \bibinfo {author} {\bibfnamefont
  {S.}~\bibnamefont {Montangero}},\ and\ \bibinfo {author} {\bibfnamefont
  {J.~F.}\ \bibnamefont {Sherson}},\ }\bibfield  {title} {\bibinfo {title}
  {{Remote optimization of an ultracold atoms experiment by experts and citizen
  scientists}},\ }\href {https://doi.org/10.1073/pnas.1716869115} {\bibfield
  {journal} {\bibinfo  {journal} {Proc. Natl. Acad. Sci. U.S.A.}\ }\textbf
  {\bibinfo {volume} {115}},\ \bibinfo {pages} {E11231} (\bibinfo {year}
  {2018})}\BibitemShut {NoStop}%
\bibitem [{\citenamefont {Khaneja}\ \emph {et~al.}(2005)\citenamefont
  {Khaneja}, \citenamefont {Reiss}, \citenamefont {Kehlet}, \citenamefont
  {Schulte-Herbr\"uggen},\ and\ \citenamefont {Glaser}}]{GRAPE}%
  \BibitemOpen
  \bibfield  {author} {\bibinfo {author} {\bibfnamefont {N.}~\bibnamefont
  {Khaneja}}, \bibinfo {author} {\bibfnamefont {T.}~\bibnamefont {Reiss}},
  \bibinfo {author} {\bibfnamefont {C.}~\bibnamefont {Kehlet}}, \bibinfo
  {author} {\bibfnamefont {T.}~\bibnamefont {Schulte-Herbr\"uggen}},\ and\
  \bibinfo {author} {\bibfnamefont {S.~J.}\ \bibnamefont {Glaser}},\ }\bibfield
   {title} {\bibinfo {title} {{Optimal control of coupled spin dynamics: design
  of {NMR} pulse sequences by gradient ascent algorithms}},\ }\href
  {https://doi.org/https://doi.org/10.1016/j.jmr.2004.11.004} {\bibfield
  {journal} {\bibinfo  {journal} {J. Magn. Reson.}\ }\textbf {\bibinfo {volume}
  {172}},\ \bibinfo {pages} {296} (\bibinfo {year} {2005})}\BibitemShut
  {NoStop}%
\bibitem [{\citenamefont {Hu}\ \emph {et~al.}(2019)\citenamefont {Hu},
  \citenamefont {Ma}, \citenamefont {Cai}, \citenamefont {Mu}, \citenamefont
  {Xu}, \citenamefont {Wang}, \citenamefont {Wu}, \citenamefont {Wang},
  \citenamefont {Song}, \citenamefont {Zou}, \citenamefont {Girvin},
  \citenamefont {Duan},\ and\ \citenamefont {Sun}}]{sunGRAPE1}%
  \BibitemOpen
  \bibfield  {author} {\bibinfo {author} {\bibfnamefont {L.}~\bibnamefont
  {Hu}}, \bibinfo {author} {\bibfnamefont {Y.}~\bibnamefont {Ma}}, \bibinfo
  {author} {\bibfnamefont {W.}~\bibnamefont {Cai}}, \bibinfo {author}
  {\bibfnamefont {X.}~\bibnamefont {Mu}}, \bibinfo {author} {\bibfnamefont
  {Y.}~\bibnamefont {Xu}}, \bibinfo {author} {\bibfnamefont {W.}~\bibnamefont
  {Wang}}, \bibinfo {author} {\bibfnamefont {Y.}~\bibnamefont {Wu}}, \bibinfo
  {author} {\bibfnamefont {H.}~\bibnamefont {Wang}}, \bibinfo {author}
  {\bibfnamefont {Y.~P.}\ \bibnamefont {Song}}, \bibinfo {author}
  {\bibfnamefont {C.-L.}\ \bibnamefont {Zou}}, \bibinfo {author} {\bibfnamefont
  {S.~M.}\ \bibnamefont {Girvin}}, \bibinfo {author} {\bibfnamefont {L.-M.}\
  \bibnamefont {Duan}},\ and\ \bibinfo {author} {\bibfnamefont
  {L.}~\bibnamefont {Sun}},\ }\bibfield  {title} {\bibinfo {title} {{Quantum
  error correction and universal gate set operation on a binomial bosonic
  logical qubit}},\ }\href {https://doi.org/10.1038/s41567-018-0414-3}
  {\bibfield  {journal} {\bibinfo  {journal} {Nat. Phys.}\ }\textbf {\bibinfo
  {volume} {15}},\ \bibinfo {pages} {503} (\bibinfo {year} {2019})}\BibitemShut
  {NoStop}%
\bibitem [{\citenamefont {Ma}\ \emph {et~al.}(2020{\natexlab{a}})\citenamefont
  {Ma}, \citenamefont {Xu}, \citenamefont {Mu}, \citenamefont {Cai},
  \citenamefont {Hu}, \citenamefont {Wang}, \citenamefont {Pan}, \citenamefont
  {Wang}, \citenamefont {Song}, \citenamefont {Zou},\ and\ \citenamefont
  {Sun}}]{sunGRAPE2}%
  \BibitemOpen
  \bibfield  {author} {\bibinfo {author} {\bibfnamefont {Y.}~\bibnamefont
  {Ma}}, \bibinfo {author} {\bibfnamefont {Y.}~\bibnamefont {Xu}}, \bibinfo
  {author} {\bibfnamefont {X.}~\bibnamefont {Mu}}, \bibinfo {author}
  {\bibfnamefont {W.}~\bibnamefont {Cai}}, \bibinfo {author} {\bibfnamefont
  {L.}~\bibnamefont {Hu}}, \bibinfo {author} {\bibfnamefont {W.}~\bibnamefont
  {Wang}}, \bibinfo {author} {\bibfnamefont {X.}~\bibnamefont {Pan}}, \bibinfo
  {author} {\bibfnamefont {H.}~\bibnamefont {Wang}}, \bibinfo {author}
  {\bibfnamefont {Y.~P.}\ \bibnamefont {Song}}, \bibinfo {author}
  {\bibfnamefont {C.-L.}\ \bibnamefont {Zou}},\ and\ \bibinfo {author}
  {\bibfnamefont {L.}~\bibnamefont {Sun}},\ }\bibfield  {title} {\bibinfo
  {title} {{Error-transparent operations on a logical qubit protected by
  quantum error correction}},\ }\href
  {https://doi.org/10.1038/s41567-020-0893-x} {\bibfield  {journal} {\bibinfo
  {journal} {Nat. Phys.}\ }\textbf {\bibinfo {volume} {16}},\ \bibinfo {pages}
  {827} (\bibinfo {year} {2020}{\natexlab{a}})}\BibitemShut {NoStop}%
\bibitem [{\citenamefont {Ma}\ \emph {et~al.}(2020{\natexlab{b}})\citenamefont
  {Ma}, \citenamefont {Pan}, \citenamefont {Cai}, \citenamefont {Mu},
  \citenamefont {Xu}, \citenamefont {Hu}, \citenamefont {Wang}, \citenamefont
  {Wang}, \citenamefont {Song}, \citenamefont {Yang}, \citenamefont {Zheng},\
  and\ \citenamefont {Sun}}]{sunGRAPE3}%
  \BibitemOpen
  \bibfield  {author} {\bibinfo {author} {\bibfnamefont {Y.}~\bibnamefont
  {Ma}}, \bibinfo {author} {\bibfnamefont {X.}~\bibnamefont {Pan}}, \bibinfo
  {author} {\bibfnamefont {W.}~\bibnamefont {Cai}}, \bibinfo {author}
  {\bibfnamefont {X.}~\bibnamefont {Mu}}, \bibinfo {author} {\bibfnamefont
  {Y.}~\bibnamefont {Xu}}, \bibinfo {author} {\bibfnamefont {L.}~\bibnamefont
  {Hu}}, \bibinfo {author} {\bibfnamefont {W.}~\bibnamefont {Wang}}, \bibinfo
  {author} {\bibfnamefont {H.}~\bibnamefont {Wang}}, \bibinfo {author}
  {\bibfnamefont {Y.~P.}\ \bibnamefont {Song}}, \bibinfo {author}
  {\bibfnamefont {Z.-B.}\ \bibnamefont {Yang}}, \bibinfo {author}
  {\bibfnamefont {S.-B.}\ \bibnamefont {Zheng}},\ and\ \bibinfo {author}
  {\bibfnamefont {L.}~\bibnamefont {Sun}},\ }\bibfield  {title} {\bibinfo
  {title} {Manipulating complex hybrid entanglement and testing multipartite
  {Bell} inequalities in a superconducting circuit},\ }\href
  {https://doi.org/10.1103/PhysRevLett.125.180503} {\bibfield  {journal}
  {\bibinfo  {journal} {Phys. Rev. Lett.}\ }\textbf {\bibinfo {volume} {125}},\
  \bibinfo {pages} {180503} (\bibinfo {year} {2020}{\natexlab{b}})}\BibitemShut
  {NoStop}%
\bibitem [{\citenamefont {Xu}\ \emph {et~al.}(2020)\citenamefont {Xu},
  \citenamefont {Ma}, \citenamefont {Cai}, \citenamefont {Mu}, \citenamefont
  {Dai}, \citenamefont {Wang}, \citenamefont {Hu}, \citenamefont {Li},
  \citenamefont {Han}, \citenamefont {Wang}, \citenamefont {Song},
  \citenamefont {Yang}, \citenamefont {Zheng},\ and\ \citenamefont
  {Sun}}]{sunGRAPE4}%
  \BibitemOpen
  \bibfield  {author} {\bibinfo {author} {\bibfnamefont {Y.}~\bibnamefont
  {Xu}}, \bibinfo {author} {\bibfnamefont {Y.}~\bibnamefont {Ma}}, \bibinfo
  {author} {\bibfnamefont {W.}~\bibnamefont {Cai}}, \bibinfo {author}
  {\bibfnamefont {X.}~\bibnamefont {Mu}}, \bibinfo {author} {\bibfnamefont
  {W.}~\bibnamefont {Dai}}, \bibinfo {author} {\bibfnamefont {W.}~\bibnamefont
  {Wang}}, \bibinfo {author} {\bibfnamefont {L.}~\bibnamefont {Hu}}, \bibinfo
  {author} {\bibfnamefont {X.}~\bibnamefont {Li}}, \bibinfo {author}
  {\bibfnamefont {J.}~\bibnamefont {Han}}, \bibinfo {author} {\bibfnamefont
  {H.}~\bibnamefont {Wang}}, \bibinfo {author} {\bibfnamefont {Y.~P.}\
  \bibnamefont {Song}}, \bibinfo {author} {\bibfnamefont {Z.-B.}\ \bibnamefont
  {Yang}}, \bibinfo {author} {\bibfnamefont {S.-B.}\ \bibnamefont {Zheng}},\
  and\ \bibinfo {author} {\bibfnamefont {L.}~\bibnamefont {Sun}},\ }\bibfield
  {title} {\bibinfo {title} {Demonstration of controlled-phase gates between
  two error-correctable photonic qubits},\ }\href
  {https://doi.org/10.1103/PhysRevLett.124.120501} {\bibfield  {journal}
  {\bibinfo  {journal} {Phys. Rev. Lett.}\ }\textbf {\bibinfo {volume} {124}},\
  \bibinfo {pages} {120501} (\bibinfo {year} {2020})}\BibitemShut {NoStop}%
\bibitem [{\citenamefont {Omran}(2022)}]{Ahmed-letter}%
  \BibitemOpen
  \bibfield  {author} {\bibinfo {author} {\bibfnamefont {A.}~\bibnamefont
  {Omran}},\ }\bibfield  {title} {\bibinfo {title} {Personal communication},\
  }\href@noop {} {\  (\bibinfo {year} {2022})}\BibitemShut {NoStop}%
\bibitem [{\citenamefont {de~L\'es\'eleuc}\ \emph {et~al.}(2018)\citenamefont
  {de~L\'es\'eleuc}, \citenamefont {Barredo}, \citenamefont {Lienhard},
  \citenamefont {Browaeys},\ and\ \citenamefont {Lahaye}}]{dopp-PRA}%
  \BibitemOpen
  \bibfield  {author} {\bibinfo {author} {\bibfnamefont {S.}~\bibnamefont
  {de~L\'es\'eleuc}}, \bibinfo {author} {\bibfnamefont {D.}~\bibnamefont
  {Barredo}}, \bibinfo {author} {\bibfnamefont {V.}~\bibnamefont {Lienhard}},
  \bibinfo {author} {\bibfnamefont {A.}~\bibnamefont {Browaeys}},\ and\
  \bibinfo {author} {\bibfnamefont {T.}~\bibnamefont {Lahaye}},\ }\bibfield
  {title} {\bibinfo {title} {{Analysis of imperfections in the coherent optical
  excitation of single atoms to Rydberg states}},\ }\href
  {https://doi.org/10.1103/PhysRevA.97.053803} {\bibfield  {journal} {\bibinfo
  {journal} {Phys. Rev. A}\ }\textbf {\bibinfo {volume} {97}},\ \bibinfo
  {pages} {053803} (\bibinfo {year} {2018})}\BibitemShut {NoStop}%
\bibitem [{\citenamefont {Wu}\ \emph {et~al.}(2021)\citenamefont {Wu},
  \citenamefont {Wang}, \citenamefont {Han}, \citenamefont {Su}, \citenamefont
  {Xia}, \citenamefont {Jiang},\ and\ \citenamefont {Song}}]{Wu-Dopp}%
  \BibitemOpen
  \bibfield  {author} {\bibinfo {author} {\bibfnamefont {J.-L.}\ \bibnamefont
  {Wu}}, \bibinfo {author} {\bibfnamefont {Y.}~\bibnamefont {Wang}}, \bibinfo
  {author} {\bibfnamefont {J.-X.}\ \bibnamefont {Han}}, \bibinfo {author}
  {\bibfnamefont {S.-L.}\ \bibnamefont {Su}}, \bibinfo {author} {\bibfnamefont
  {Y.}~\bibnamefont {Xia}}, \bibinfo {author} {\bibfnamefont {Y.}~\bibnamefont
  {Jiang}},\ and\ \bibinfo {author} {\bibfnamefont {J.}~\bibnamefont {Song}},\
  }\bibfield  {title} {\bibinfo {title} {{Resilient quantum gates on
  periodically driven Rydberg atoms}},\ }\href
  {https://doi.org/10.1103/PhysRevA.103.012601} {\bibfield  {journal} {\bibinfo
   {journal} {Phys. Rev. A}\ }\textbf {\bibinfo {volume} {103}},\ \bibinfo
  {pages} {012601} (\bibinfo {year} {2021})}\BibitemShut {NoStop}%
\bibitem [{\citenamefont {Beterov}\ \emph {et~al.}(2009)\citenamefont
  {Beterov}, \citenamefont {Ryabtsev}, \citenamefont {Tretyakov},\ and\
  \citenamefont {Entin}}]{Rblifetime}%
  \BibitemOpen
  \bibfield  {author} {\bibinfo {author} {\bibfnamefont {I.~I.}\ \bibnamefont
  {Beterov}}, \bibinfo {author} {\bibfnamefont {I.~I.}\ \bibnamefont
  {Ryabtsev}}, \bibinfo {author} {\bibfnamefont {D.~B.}\ \bibnamefont
  {Tretyakov}},\ and\ \bibinfo {author} {\bibfnamefont {V.~M.}\ \bibnamefont
  {Entin}},\ }\bibfield  {title} {\bibinfo {title} {{Quasiclassical
  calculations of blackbody-radiation-induced depopulation rates and effective
  lifetimes of Rydberg $nS$, $nP$, and $nD$ alkali-metal atoms with
  $n\ensuremath{\le}80$}},\ }\href {https://doi.org/10.1103/PhysRevA.79.052504}
  {\bibfield  {journal} {\bibinfo  {journal} {Phys. Rev. A}\ }\textbf {\bibinfo
  {volume} {79}},\ \bibinfo {pages} {052504} (\bibinfo {year}
  {2009})}\BibitemShut {NoStop}%
\end{thebibliography}%
\end{document}